\newcommand{\be}{\begin{equation}}
\newcommand{\ee}{\end{equation}}
\newcommand{\bea}{\begin{eqnarray}}
\newcommand{\eea}{\end{eqnarray}}

\def\pd{{\partial}}
\newcommand{\md}{{\rm{d}}}

    \newcommand{\lfd}[1]{\displaystyle \frac{\mathrm{D} #1}{\mathrm{D} t}}

\documentclass[intlimits,twoside,a4paper]{article}
\usepackage{graphicx}
\usepackage[T2A]{fontenc}
\usepackage[cp1251]{inputenc}
\usepackage{amsmath}
\usepackage{amssymb}
\usepackage{bm}
\usepackage[eqsecnum]{cmpj}

\issue{2011}{14}{1}{13401}
\doinumber{10.5488/CMP.14.13401}

\title[Constitutive equations for granular flow]%
{Constitutive equations for granular flow with uniform
 mean shear and spin fields
}
  \author{K. Takechi, K. Yoshida, T. Arimitsu}
  \address{Graduate School of Pure and Applied Sciences, University of Tsukuba, \\1--1--1 Tennodai, Ibaraki 305--8571, Japan}

\date{Received May 12, 2010, in final form December 22, 2010}
\begin{document}

\maketitle

\begin{abstract}
Numerical simulations of two-dimensional granular flows under
uniform shear and external body torque were performed in order to
extract the constitutive equations for the system. The outcome of
the numerical simulations is analyzed on the basis of the
micropolar fluid model. Uniform mean shear field and mean spin
field, which is not subordinate to the vorticity field, are
realized in the simulations. The estimates of stresses based on
kinetic theory by Lun [Lun, J. Fluid Mech., 1991, {\bf 233}, 539]
are in good agreement with the simulation results for a low area
fraction $\nu=0.1$ but the agreement becomes weaker as the area
fraction gets higher.  However, the estimates in the kinetic
theory can be fitted to the simulation results up to $\nu=0.7$ by
renormalizing the coefficient of roughness.  For a relatively
dense granular flow ($\nu=0.8$), the simulation results are also
compared with Kanatani's theory [Kanatani, Int. J. Eng. Sci.,
1979, {\bf 17}, 419]. It is found that the dissipation function
and its decomposition into the constitutive equations in
Kanatani's theory are not consistent with the simulation results.

\keywords granular flow, constitutive equation, micropolar fluid, kinetic equation
\pacs 45.70.Mg, 47.57.Gc, 47.50.-d
\end{abstract}

\section{Introduction}

Collective motions of granular materials behave like fluid motions
under appropriate conditions.  However, unlike the Newtonian fluids, the
basic equations for the collective motions of granular materials
have not been well established yet.  One of the difficulties of the
problem lies in the fact that the scale of macroscopic collective motions
is not well separated from the microscopic scale of
the system such as the radius of the granular particles.
Thus, applicability of arguments based on the scale separation
would be limited.  Many detailed properties of the individual
particles would  directly affect the behavior of the macroscopic
flow.

One possible way to treat such granular flows is to model them as
flows of a micropolar fluid, a fluid with polar micro-structures
such as spin~\cite{CondiffDahler1964,Eringen1966}. By applying
micropolar fluid mechanics to granular flows, the spin of the
granular particles can be coupled to the dynamics of the
macroscopic collective motions of the granular particles.  The
microscopic properties of the granular particles are reflected in
the equations of motion for the macroscopic fields through the
constitutive equations, i.e., the relations between strains and
stresses (see section~\ref{sec:micropolar}).

For sparse and rapid granular flows, the equations of motion as a
micropolar fluid can be derived within the framework of a kinetic
theory.  Firstly, the kinetic theory was developed without
introducing the frictional interactions between particles and the
degrees of freedom for spin by Savage and
Jeffrey~\cite{SavageJeffrey1981}, and Jenkins and
Savage~\cite{JenkinsSavage1983}. Although Jenkins and
Richman~\cite{JenkinsRichman1985}, Jenkins and
Zhang~\cite{JenkinsZhang2002} and Yoon and
Jenkins~\cite{YoonJenkins2005} introduced frictional interaction
between particles to the kinetic theory, they eliminated the
macroscopic degrees of freedom of the spin field by assuming that
the macroscopic spin field is subordinate to the vorticity field.
Such an assumption may be justified, for example, for steady flows
far from the boundary.
In the kinetic theories, the effect of frictional interactions can
be absorbed into the renormalized restitution coefficient. Saitoh
and Hayakawa~\cite{SaitohHayakawa2007} performed numerical
simulations of two-dimensional granular flow under a plane shear
and confirmed that the hydrodynamic equations derived from the
kinetic theories agree with the simulation results. In some cases,
such as flows near boundaries, discrepancy between the spin field
and the vorticity field is not negligible.  The kinetic theories
retaining the spin field as
 independent macroscopic degrees of freedom were developed
by Lun and Savage~\cite{LunSavage1987}, Lun~\cite{Lun1991} and
Goldshtein and Shapiro~\cite{GoldshteinShapiro1995}. Mitarai et
al.~\cite{MitaraiHayakawaNakanishi2002} performed numerical
simulation of a collisional granular flow on a slope and showed
that the velocity and spin field profiles are in agreement with
the micropolar fluid equations based on constitutive equations
which are consistent with that in~\cite{Lun1991}.

When the granular particles become dense enough and the volume
fraction exceeds the critical value, the collective motions of
particles stop to behave like a fluid in a sense that a finite
shear stress is required to create an infinitesimal strain.  Such
a phase is called the jammed phase. The phase transition between
unjammed phase and jammed phase is called the jamming transition.
Scaling laws near the critical point of the jamming transition
have been suggested and verified in the numerical simulations by
Hatano~\cite{Hatano2008}, and Otsuki and
Hayakawa~\cite{OtsukiHayakawa2009,OtsukiHayakawa2009PRE}. The
frictional interactions among particles were not considered in
their studies. Recently, a number of results on the jamming
transition based on numerical simulations including the frictional
interactions have been reported (e.g., Silbert et
al.~\cite{Silbert2002}, Zhang and Makse~\cite{ZhangMakse2005} and
Shundyak, Hecke and Saarloos~\cite{ShundyakHeckeSaarloos2007}.)

There can be a substantial intermediate regime of
the volume fraction
between the kinetic region with low volume fraction
and the critical region near the jamming transition.
In this regime,  interaction of $n$-particles with $n>2$ would
become important. Kanatani~\cite{Kanatani1979int} developed a
micropolar fluid theory for relatively dense granular flows in
which particles are almost regularly in contact with the other
particles.  The regime where Kanatani's theory is applicable is
possibly located in this intermediate regime. Kano et
al.~\cite{KanoShimosakaHidaka1996} showed that numerical
simulation of a granular flow on an inclined trough is in
qualitative agreement with the micropolar fluid equation based on
Kanatani's theory regarding the velocity profile.

In this paper, we focus on the constitutive equations for granular flows.
As an intrinsic nature of the granular flows, the spin field associated with
granular particles is not subordinate to the velocity field of their mean flow.
This situation is analogous to the case of micropolar fluids in which
the spin field is regarded as an independent degree of freedom.
As we will see in section~\ref{sec:micropolar},
both the difference between the vorticity and spin fields,
which will be denoted by $R_{ji}$,
and the spatial derivative of the spin field,
which will be denoted by $\Omega_{kji}$,
contribute to the constitutive equations.
From a theoretical point of view,
it is desirable to analyze them separately.
Therefore, let us consider the case of $R_{ji}\ne 0$ and
$\Omega_{kji}=0$.  Note that $R_{ji}\ne 0$ near the boundary.
When sufficient numbers of particles are contained in a region near
the boundary with the length scale smaller than the typical length scale
in which the shear and spin fields changes, we may consider
that uniform shear and spin fields (i.e. $\Omega_{kji}=0$) with
$R_{ji}\ne 0$ are approximately realized in the region.
The situation $R_{ji}\ne 0$ and $\Omega_{kji}=0$ would be also
obtained by applying external torque to each particle.
Note that, provided that the micropolar fluid
picture is appropriate for the granular flows,
the constitutive equations depend solely on velocity,
spin fields and their spatial derivatives at the local point
under consideration and independent of driving forces that generate
the fields.
A possible way to apply the external torque to each particle
in experiments is to embed the source of angular momentum inside
each particle.  That is, the particle is supposed to be a kind of
micro-machine composed of an outer shell
and an inner sphere with the friction between them being small.
Initially, the inner sphere is made to rotate with a high
angular velocity by some means while the outer shell is not rotating.
Then, the angular momentum of the inner sphere is continuously
supplied to the outer shell through the friction until
the inner sphere loses its substantial angular momentum.
By virtue of small friction, one can realize a longer period
for the experiment.
By considering the inner sphere as an exterior system,
the situation implies that the external torque is continuously
applied to the particle (the outer shell).
The inner sphere can be replaced by a liquid with low viscosity
such as a super fluid.
The actual setting of the above system
for the experiment may be quite difficult.  However, in numerical
experiments, it is quite easy to apply the external torque to each
particle.

Taking the above into consideration, we performed numerical
simulations of two-dimensional granular flows under uniform shear
and uniform external torque field. By virtue of the external
torque field and the applied boundary conditions, macroscopically
uniform vorticity and spin fields are realized and their
magnitudes are controlled independently, which means that
$\Omega_{kji}=0$ and the magnitude of $R_{ji}$ can be controlled
(see section~\ref{sec:simulation}). Thus, we concentrate on the
$R_{ji}$ dependence of the constitutive equations with
$\Omega_{kji}$ fixed to $0$.  The study of the $\Omega_{kji}$
dependence of the constitutive equations will be the next step and
will not be referred to in this paper.  Unlike the preceding
numerical studies such as~\cite{KanoShimosakaHidaka1996}
and~\cite{MitaraiHayakawaNakanishi2002}, we are able to obtain not
only the velocity and spin field profiles, which are the results
of the constitutive equations, but also the constitutive equations
directly. Since the subject of this paper is the micropolar fluid
aspect of the granular flows, the value of area fraction is varied
within the unjammed region.  We compare the results from the
numerical simulations with those from the kinetic theory by Lun
\cite{Lun1991}, which is capable of treating cases that the spin
field is not subordinate to the vorticity field.  For the
intermediate regime noted above, we also compared the simulation
results with Kanatani's theory~\cite{Kanatani1979int}.

This paper is organized as follows.  In
section~\ref{sec:micropolar}, a brief review of the micropolar
fluid theory is given. In sections~\ref{sec:kinetic}
and~\ref{sec:Kanatani}, the kinetic theory by Lun and Kanatani's
theory are reviewed, respectively. In section~\ref{sec:comments},
comments on the two theories are given.  In
section~\ref{sec:simulation}, the results of the numerical
simulations are shown and they are compared with the theories.  In
section~\ref{sec:discussion}, discussion is presented.

\section{Equations for micropolar fluid}
\label{sec:micropolar}

In this paper, we consider collective motions of particles.
For simplicity, we assume that the particles are of the
same mass $m$ and the same moment of inertia $I$.
Let $c_i(t), w_{ji}(t)$ and $r_i(t)$  be, respectively,
the velocity, the spin and the position of the particle at time
$t$ where $i$ and $j$ are coordinate indices of $d$-dimensional space.
Here, $d$ can formally be an arbitrary positive integer larger than $1$.
In this paper, we use the convention that the spin is
expressed by a skew-symmetric tensor $w_{ji}$
whose $(j,i)$-th component gives the angular velocity in
the $(j,i)$ coordinate plane.
Let $F^{(N)}({\bm c}^{(1)},{\bm w}^{(1)},{\bm r}^{(1)};\cdots;
{\bm c}^{(N)},{\bm w}^{(N)},{\bm r}^{(N)};t)$ be the probability
density function in the phase space of $N$-particles system
satisfying Liouville equation.
Here, the bold letters $\bm c$, $\bm w$ and  $\bm r$ denote vector or
tensor, the superscript $(\alpha)$ on ${\bm c}^{(\alpha)},
{\bm w}^{(\alpha)}$ and ${\bm r}^{(\alpha)}$ is the index of the
particle and $F^{(N)}$ is symmetrized with respect to interchanges
of the particles.
The $s$-particles set distribution function $f^{(s)} (s \leqslant N)$ is
given by
\begin{align}
f^{(s)}({\bm c}^{(1)},{\bm w}^{(1)},{\bm r}^{(1)};\cdots;
{\bm c}^{(s)},{\bm w}^{(s)},{\bm r}^{(s)};t)%
=&\frac{N!}{(N-s)!}
\prod_{\alpha=s+1}^{N}\left(\int\md{\bm c}^{(\alpha)}
\int \md{\bm w}^{(\alpha)} \int\md{\bm r}^{(\alpha)}\right)\nonumber\\[1ex]
&{} \times F^{(N)}({\bm c}^{(1)},{\bm w}^{(1)},{\bm r}^{(1)};\cdots;
{\bm c}^{(N)},{\bm w}^{(N)},{\bm r}^{(N)};t),
\end{align}
and the number density of the $s$-particles sets
$n^{(s)}$ is given by
\begin{align}
n^{(s)}({\bm r}^{(1)},\cdots,{\bm r}^{(s)};t)=&
\prod_{\alpha=1}^s \left(\int \md {\bm c}^{(\alpha)}
\int\md {\bm w}^{(\alpha)}\right)\nonumber\\[1ex]
&{}\times
f^{(s)}({\bm c}^{(1)},{\bm w}^{(1)},{\bm r}^{(1)};\cdots;
{\bm c}^{(s)},{\bm w}^{(s)},{\bm r}^{(s)};t).
\end{align}
Macroscopic fields such as the mass density field $\rho({\bm r},t)$,
the moment of inertia density field $\rho_I({\bm r},t)$,
the velocity field ${\bm v}({\bm r},t)$ and
the spin field ${\bm \omega}({\bm r},t)$
are introduced as
\begin{align}
&\rho({\bm r},t):=m n^{(1)}({\bm r},t),
&&\rho_I({\bm r},t):=I n^{(1)}({\bm r},t),\\
&{\bm v}({\bm r},t):=\langle {\bm c}\rangle_{\bm r,t}\,,
&&{\bm \omega}({\bm r},t):=\langle {\bm w}\rangle_{\bm r,t}\,,
\end{align}
where
\be
\langle \psi({\bm c},{\bm w}) \rangle_{{\bm r},t}:=
\frac{1}{n^{(1)}({\bm r},t)}
\int \md{\bm c} \int \md{\bm w}
\psi({\bm c},{\bm w})
f^{(1)}({\bm c},{\bm w},{\bm r};t),
\ee
for an arbitrary function $\psi$ of ${\bm c}$ and ${\bm w}$.
Hereafter, indices or subscripts of spatial or time coordinates
will be suppressed unless we need to emphasize them.

These macroscopic fields satisfy the following equations,
\begin{align}
\lfd{\rho} &+ \rho \partial_{i} v_{i} = 0,
\label{eqn.continuity} \\
\lfd{\rho_{I}} &+ \rho_{I} \partial_{i} v_{i} = 0,
\label{eqn.inertia_moment_continuity} \\
\rho \lfd{v_{i}} &= \partial_{j} \sigma_{ji} + \rho b_{i}\,,
 \label{eqn.motion} \\
\rho_{I} \lfd{\omega_{ji}} &= 2\sigma_{[ji]} + \partial_{k} \lambda_{kji} +
\rho_I \tau_{ji}\,,\label{eqn.anglar_motion}
\end{align}
where
$D/Dt = \pd/\pd t + v_i \pd_i$,
$\sigma_{ji}$ is the stress tensor,
$\lambda_{kji}$ the couple-stress tensor,
$b_i$ the external body force and $\tau_{ji}$ the external body torque.
Here and hereafter, summations are taken over repeated subscripts
and we employ the notations,
\begin{gather}
T_{(ji)}=\frac{1}{2}\left( T_{ji} + T_{ij}\right), \quad
T_{[ji]}=\frac{1}{2}\left( T_{ji} - T_{ij}\right), \\
T_{\{ji\}}=T_{(ji)}-\frac{1}{d}T_{kk}\delta_{ji}\,,
\end{gather}
for an arbitrary tensor $T_{ji}$.
Equations (\ref{eqn.continuity})--(\ref{eqn.anglar_motion})
correspond, respectively, to the conservation laws of mass,
moment of inertia, momentum and angular momentum.
Note that $v_i$ and $\omega_{ji}$ are mutually independent
degrees of freedom.
When there is no spin field $\omega_{ji}$,
(\ref{eqn.continuity})--(\ref{eqn.motion}) reduce
to the equations of ordinary fluids.  A fluid with
the degree of freedom of the spin field is called micropolar
fluid.  The equations of motions will be closed when the
stresses $\sigma_{ji}$ and $\lambda_{kji}$ are known in terms of
$\pd_j v_i$ and $\Omega_{kji}:=\pd_k \omega_{ji}$.
The equations which yield such relations are called
 constitutive equations.

Let $\bm C$ 
and $\bm W$ 
be the fluctuations of
velocity $\bm c$ and spin $\bm w$ of individual particles
around the macroscopic fields, i.e.,
\be
\bm C=\bm c - \bm v,\qquad
\bm W=\bm w - \bm \omega.
\label{eq:defCW}
\ee
The ``internal energies'' per unit mass
$\epsilon_{\rm t}$ and $\epsilon_{\rm r}$ associated with
the fluctuations $\bm C$ and $\bm W$, respectively, are given by
\be
\epsilon_{\rm t}:= \frac{1}{2}
\Bigl\langle C_i C_i \Bigr\rangle,\qquad
\epsilon_{\rm r}:= \frac{\rho_I}{4\rho}
\Bigl\langle W_{ji} W_{ji} \Bigr\rangle.
\ee
Here, the subscripts t and r indicate the translational and
rotational motions, respectively.
Two kinds of ``temperature'', $T_{\rm t}$ and $T_{\rm r}$, are
introduced by the relations $\epsilon_{\rm t}=(d/2)T_{\rm t}$ and
$\epsilon_{\rm r}=(d(d-1)/4)T_{\rm r}$.
The total ``internal energy'' is given by
$\epsilon_{\rm U}=\epsilon_{\rm t}+\epsilon_{\rm r}$.
Here, we have assumed that the particles are rigid bodies and that there
is no potential force acting among particles, that is, there is neither
contribution of elastic energy nor of potential energy to
the ``internal energy''.
One can show from (\ref{eqn.continuity})--(\ref{eqn.anglar_motion})
that the kinetic energy of macroscopic fields per unit mass,
\be
\epsilon_{\rm K}=\frac{1}{2} v_i v_i
+ \frac{\rho_I}{4\rho}
\omega_{ji}\omega_{ji}\,,
\ee
satisfies
\be
\rho\frac{D\epsilon_{\rm K}}{D t} = \Psi + p \pd_i v_i - \Phi,
\label{eq:dekdt}
\ee
with
\begin{align}
\Psi &:= \pd_k \left(\sigma_{ki}v_i\right)
+ \frac{1}{2}\pd_k\left(\lambda_{kji}\omega_{ji}\right)
+\rho b_i v_i + \frac{1}{2}\rho_I \tau_{ji}\omega_{ji}\,, \\
\Phi &:= 
\sigma_{\{ji\}} E_{ji}+\sigma_{[ji]} R_{ji}
+\frac{1}{2} \lambda_{kji} \Omega_{kji}\,,
\label{eqn.Phi_granular}
\end{align}
where we have introduced notations,
\be
E_{ji}:=\pd_{\{j} v_{i\}},\qquad R_{ji}:=\pd_{[j} v_{i]}-\omega_{ji}\,,
\label{eq:EROmega}
\ee
and $p=-\sigma_{ii}/d$ is the pressure.
Note that $\Psi$ is the work done on a fluid element
per unit volume per unit time by the stress $\sigma_{ji}$,
the couple stress $\lambda_{kji}$, the external body force $b_i$
and the external body torque $\tau_{ji}$.
From (\ref{eq:dekdt}) and the energy budget equation,
\be
\rho\frac{D}{Dt}(\epsilon_{\rm K}+\epsilon_{\rm U}) = \Psi + q -\pd_i h_i\,,
\ee
we obtain
\be
\rho \frac{D\epsilon_{\rm U}}{Dt}=\Phi - p \pd_i v_i + q - \partial_{i} h_{i}\,,
\label{eqn.energy_flow}
\ee
where $q$ is input of the ``internal energy'' per unit volume
per unit time, $h_j$ is the ``internal energy'' flux.
The quantity $\Phi- p \pd_iv_i$ gives the energy transferred from
the kinetic energy to the ``internal energy'' per unit volume
per unit time.  When macroscopic fields $\rho, \pd_j v_i, \epsilon_U$
and $h_j$ are constant in time and space,
we have $\pd_i v_i=0$ from (\ref{eqn.continuity}), and
(\ref{eqn.energy_flow}) reduces to
\be
\Phi=-q.
\label{eq:phiq}
\ee
The equation (\ref{eq:phiq}) implies that the energy transfer rate
$\Phi$ from the kinetic energy to the ``internal energy'' balances
with the dissipation rate of ``internal energy'' $-q$ for
macroscopically steady states.  $\Phi$ is called the dissipation function
since it gives the energy going out from the kinetic
energy $\epsilon_{\rm K}$ per unit time per unit volume
in the macroscopically steady state.

\section{Kinetic theory for collisional granular flow}
\label{sec:kinetic} A Kinetic theory for collisional granular flow
is developed by Savage and Jeffrey~\cite{SavageJeffrey1981}, and
Jenkins and Savage~\cite{JenkinsSavage1983}.  It is extended to
include the effects of surface friction and inertial moment of
particles by Lun and Savage~\cite{LunSavage1987}, and Lun
\cite{Lun1991}.  From assumptions on the collisional process of
two particles, the constitutive equations for the granular
material as a micropolar fluid are derived. We briefly review the
theory by Lun~\cite{Lun1991} in the following.

\begin{figure}
\centerline{\includegraphics[width=0.3\linewidth]{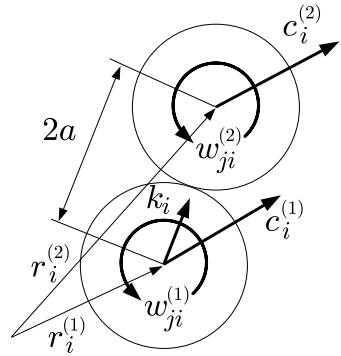}}
\caption{\label{fig:2_grains} Configuration of two contacting granular particles.}
\end{figure}
The dimension of the configuration space is $d$ and
the particles are assumed to be $d$-dimensional spheres with
the same radius $a$.  The mass and the moment of inertia
are, respectively denoted by $m$ and $I$ as in section~\ref{sec:micropolar}.
Let the colliding two particles be labeled as $1$ and $2$, and a
unit vector $\bm k$ be defined by \be \bm k:=\frac{{\bm
r}^{(2)}-{\bm r}^{(1)}}{|{\bm r}^{(2)}-{\bm r}^{(1)}|}\,,
\label{eq:defk} \ee (see figure~\ref{fig:2_grains}). In this
section, we employ the notation that the quantities without (with)
check~$\check\ $ denote those just before (after) the contact.
Let $\bm J$ be the impulse of the force exerted on the particle $2$
by the particle $1$ through the contact point.  We have
\begin{gather}
m \check{c}_i^{(1)}=m c_i^{(1)}-J_i\,, \qquad
\label{eq:J1}
m \check{c}_i^{(2)}=m c_i^{(2)}+J_i\,,\\
I \check{w}_{ji}^{(1)}=
I w_{ji}^{(1)} - %
a (k_j J_i - k_i J_j),\qquad
I \check{w}_{ji}^{(2)}=
I w_{ji}^{(2)} - %
a(k_j J_i - k_i J_j).
\label{eq:J2}
\end{gather}
Let us assume that the velocity difference $\xi_i$ between the
two surfaces at the contact point,
\be
\xi_i := (c_l^- + 
a k_{j} w_{jl}^{+})
(\delta_{li} - k_{l} k_{i}),
\label{eq:xi}
\ee
where
\be
{\bm c}^{\pm}:={\bm c}^{(1)}\pm{\bm c}^{(2)},\qquad
{\bm w}^{\pm}:={\bm w}^{(1)}\pm{\bm \omega}^{(2)},
\ee
changes as,
\begin{align}
k_i \check{\xi}_i&= -e k_i \xi_i\,,
\label{eq:e_def}\\
(\delta_{ij}-k_i k_j) \check{\xi}_j&=
-\beta (\delta_{ij} - k_i k_j) \xi_j\,,
\label{eq:beta_def}
\end{align}
after the collision.
Here, the coefficient of restitution $e$
and the coefficient of roughness $\beta$ are assumed to be
constants satisfying $0 \leqslant e \leqslant 1$ and
$-1 \leqslant \beta \leqslant 1$.
From (\ref{eq:J1}), (\ref{eq:J2}),
(\ref{eq:e_def}), (\ref{eq:beta_def}), we obtain
\begin{align}
\check{c}_i^{(1)}-c_i^{(1)}&=-\check{c}_i^{(2)}+c_i^{(2)}%
=-\eta_2 v_i^--(\eta_1 -\eta_2) k_j c_j^- k_i
-\eta_2 a k_j w_{ji}^+\,,\\
\check{w}_{ji}^{(1)}-w_{ji}^{(1)}&=
\check{w}_{ji}^{(2)}-w_{ji}^{(2)}%
=-\frac{\eta_2}{K a}\left[(k_jc_i^{-}-k_ic_j^{-}+
a (k_lk_j w_{li}^+ - k_lk_i w_{lj}^+)
\right],
\end{align}
where $\eta_1=(1+e)/2$, $\eta_2=(1+\beta) K/ 2 (K+1)$ and $K=I/ma ^2$.

Assuming the binary collisions, the equation of motion
for $\langle \psi \rangle$, where $\psi$ is an arbitrary function
of $\bm c$ and $\bm w$, can be written as
\begin{align}
\frac{\pd}{\pd t}\left(n^{(1)}\langle \psi\rangle\right)
&=-\pd_i\left( n^{(1)}\langle c_i \psi\rangle\right)
-\pd_i \theta_i(\psi) +\chi(\psi)%
+n^{(1)}
\left(\left\langle \frac{\pd \psi}{\pd c_i}\right\rangle b_i
+\frac{1}{2}\left\langle \frac{\pd \psi}{\pd w_{ji}}\right\rangle
\tau_{ji}\right),
\label{eq:psi}
\end{align}
where
\begin{align}
\theta_i(\psi;\bm r,t):=&
-\frac{1}{2}\int \md \bm c^{(1)}\int \md \bm c^{(2)}
\int\md\bm w^{(1)}\int\md\bm w^{(2)}\int_{\bm k\cdot\bm c^->0}\md \bm k\nonumber\\
&\quad\times\bm k\cdot \bm c^- (2 a)^d k_i
(\check{\psi}^{(1)}-\psi^{(1)})\nonumber\\
&\quad\times f^{(2)}\left(\bm c^{(1)},\bm w^{(1)},\bm r-a \bm k;
\bm c^{(2)},\bm w^{(2)},\bm r+a \bm k;t\right),
\label{eq:theta}\\
\chi(\psi;\bm r,t):=&
\frac{1}{2}\int \md \bm c^{(1)}\int \md \bm c^{(2)}
\int\md\bm w^{(1)}\int\md\bm w^{(2)}\int_{\bm k\cdot\bm c^->0}\md \bm k\nonumber\\
&\quad\times\bm k\cdot \bm c^- (2a)^{d-1}
(\check{\psi}^{(1)}-\psi^{(1)}+\check{\psi}^{(2)}-\psi^{(2)})\nonumber\\
&\quad\times f^{(2)}\left(\bm c^{(1)},\bm w^{(1)},\bm r-a \bm k;
\bm c^{(2)},\bm w^{(2)},\bm r+a \bm k;t\right).
\label{eq:chi}
\end{align}
By substituting $\psi=m$ and $\psi=m c_i$ in (\ref{eq:psi}), one finds that
the stress tensor $\sigma_{ji}$ can be written as
\be
\sigma_{ji}=\sigma_{ji}^{(\rm k)}+
\sigma_{ji}^{(\rm c)},
\ee
with
\begin{align}
\sigma_{ji}^{(\rm k)}:=&-\rho
\langle C_j C_i \rangle,
\label{eq:sigmak}\\
\sigma_{ji}^{(\rm c)}:=&
-\theta_j(m C_i),
\label{eq:sigmac}
\end{align}
where $\sigma_{ji}^{(\rm k)}$ denotes the kinetic contribution to the
stress $\sigma_{ji}$ due to the particles that cross the plane perpendicular
to $j$-axis, and $\sigma_{ji}^{(\rm c)}$ denotes the collisional contribution
due to the collisions of the two particles in different sides of
the plane.

The particle-pair distribution function $f^{(2)}$ is assumed to be
approximated by the form
\begin{align}
  &f^{(2)}({\bm c}^{(1)},{\bm w}^{(1)},{\bm r}-a{\bm k},
  {\bm c}^{(2)},{\bm w}^{(2)},{\bm r}+a{\bm k};t)\nonumber\\
  &\hspace{15mm}\simeq g_0(2 a;\bm r,t)
  f^{(1)}({\bm c}^{(1)},{\bm w}^{(1)},{\bm r}-a {\bm k};t)
  f^{(1)}({\bm c}^{(2)},{\bm w}^{(2)},
  {\bm r}+a {\bm k};t),
  \label{eq:f2gf1f1}
\end{align}
where $g_0(r';\bm r,t)$ is a radial distribution given by
\be
g_0(r';\bm r,t) :=
\frac{n^{(2)}({\bm r}-\frac{r'}{2}\bm e,
  {\bm r}+\frac{r'}{2}{\bm e};t)}{[n^{(1)}({\bm r},t)]^2}\,,
\label{eq:defg0}
\ee
with an assumption that it is insensitive to the direction of a unit vector
$\bm e$.
Assuming that $a$ is sufficiently smaller than the
typical spatial scale in which the amplitude of $f^{(1)}$ varies,
we have
\be
f^{(1)}({\bm c},{\bm w},{\bm r}\pm a {\bm k};t)
\simeq f^{(1)}({\bm c},{\bm w};{\bm r},t)
\left[ 1 \pm
  a {\bm k}\cdot{\bm \nabla}
  \ln f^{(1)}({\bm c},{\bm w},{\bm r};t)\right].
\ee
Let  $f^{(1)}$ be written as
\be
f^{(1)}(\bm c, \bm w, \bm r; t)=f_0^{(1)}(\bm c, \bm w, \bm r; t)
[1+\phi(\bm c, \bm w, \bm r; t)],
\label{eq:f(1)}
\ee
where $f_0^{(1)}$ is the distribution function at a local equilibrium
state and $\phi$ represents the perturbation.  The function $f_0^{(1)}$
is given by
\begin{align}
  f_0^{(1)}({\bm c},{\bm w},{\bm r};t)&=
  \frac{n^{(1)}(\bm r,t)}{(2\pi T_{\rm t}({\bm r},t))^{\frac{d}{2}}
    (2\pi m T_{\rm r}({\bm r},t)/I)^{\frac{d(d-1)}{4}}}
  \nonumber\\
  &\quad\times \exp \left[
    -\frac{C_i({\bm r},t)C_i({\bm r},t)}{2T_{\rm t}({\bm r},t)}
    -\frac{I W_{ji}({\bm r},t) W_{ji}({\bm r},t)}{4 m T_{\rm r}({\bm r},t)}
  \right].
  \label{eq:f(1)0}
\end{align}
Note that different temperatures $T_{\rm t}$ and $T_{\rm r}$ are assigned to
transitional and rotational degrees of freedom, respectively.

Assuming that the mean
is much smaller than the typical magnitude of the fluctuation
for the distribution of $w_{ji}$, i.e.,
$|\omega_{ji}| \ll \sqrt{(m/I) T_{\rm r}}$,
we have
\begin{align}
  &\exp \left\{
    -\frac{I W_{ji}({\bm r},t)W_{ji}({\bm r},t)}{4 m T_{\rm r}({\bm r},t)}\right\}
  \simeq
  \left[
    1 + \frac{I  w_{ji}\omega_{ji}({\bm r},t)}{2 m T_{\rm r}(\bm r,t)}
  \right]
  \exp \left[
    -\frac{I w_{ji} w_{ji}}{4 m T_{\rm r}({\bm r},t)}
  \right]
  .
  \label{eq:expIww}
\end{align}
It is assumed that the function $\phi$ is approximated by
a linear function of degrees of nonequilibrium such as
the symmetric and traceless kinetic stress tensor
$\sigma_{\{ij\}}^{(\rm k)}$, and the translational and
rotational kinetic energy fluxes $h_{{\rm t},i}^{(\rm k)}$ and
$h_{{\rm r},i}^{(\rm k)}$ given by
\begin{gather}
  h_{{\rm t},i}^{(\rm k)}(\bm r,t)
  :=\frac{\rho}{2}\langle C_i C_j C_j\rangle_{\bm r,t}\,,\qquad
  h_{{\rm r},i}^{(\rm k)}(\bm r,t)
  :=\frac{\rho_I}{4}\langle C_i W_{lj}W_{lj}\rangle_{\bm r,t}\,.
\end{gather}
In what follows, we restrict ourselves to the case of
macroscopic fields $\pd_j v_i, \omega_{ji}, T_{\rm t}$ and
$T_{\rm r}$ being constant in space and time.  In such a case,
there are no energy fluxes $h_{{\rm t},i}^{(\rm k)}$ and
$h_{{\rm r},i}^{(\rm k)}$ and the perturbation $\phi$
depends solely on $\sigma_{\{ij\}}^{(\rm k)}$. Let us assume the form
\be
\phi=-\frac{1}{2\rho T_{\rm t}}
\sigma_{\{ij\}}^{(\rm k)} C_i C_j\,,
\label{eq:phi}
\ee
where the factor $-1/2\rho T_{\rm t}$ is determined
from the consistency condition that (\ref{eq:sigmak}) is satisfied.
Note that the total $\sigma^{(\rm k)}$ is given by
\be
\sigma_{ji}^{(\rm k)}=-\rho T_{\rm t}\delta_{ji}+\sigma_{\{ji\}}^{(\rm k)}\,,
\ee
where there is no anti-symmetric part $\sigma_{[ji]}^{(\rm k)}$ in
the definition (\ref{eq:sigmak}).
By substituting $\psi=m C_iC_j/2$ into (\ref{eq:psi}), we obtain
\be
0=\frac{1}{2}\left(
  \sigma_{li}^{(\rm k)}\pd_l v_j + \sigma_{lj}^{(\rm k)}\pd_l v_i
\right)
+\frac{1}{2}\left(
  \sigma_{li}^{(\rm c)}\pd_l v_j + \sigma_{lj}^{(\rm c)}\pd_l v_i
\right)
+\chi\left(\frac{mC_iC_j}{2}\right),
\label{eq:psimcicj}
\ee
where $\sigma^{(\rm k)}$ and $\sigma^{(\rm c)}$ are given by
(\ref{eq:sigmak}) and (\ref{eq:sigmac}), respectively, and
$\chi(\cdot)$ by (\ref{eq:chi}).
$\sigma^{(\rm c)}$ and $\chi(mC_iC_j/2)$ reduce to
functions of $\pd_j v_i$, $\omega_{ji}$ and $\sigma^{(\rm k)}$,
after performing the integrations with respect to
$\bm k, \bm c^{(i)}$ and $\bm w^{(i)} (i=1,2)$ in (\ref{eq:theta}) and
(\ref{eq:chi}).  Substitute the functions into (\ref{eq:psimcicj})
and assume that the magnitudes of $\pd_j v_i$ and $\omega_{ji}$ are so small
that the second or higher order terms in $\pd_j v_i$ and $\omega_{ji}$
can be neglected.
Then we arrive at the equations,
\begin{align}
  \sigma_{\{ji\}}^{(\rm k)}&=2 \mu_k E_{ji}\,,
  \label{eq:sigmasymk}\\
  \sigma_{\{ji\}}^{(\rm c)}&=
  \frac{2^{d}}{d+2}
  g_0\nu \left[(4 \eta_1 + d\eta_2) \rho a
    \left(\frac{T_{\rm t}}{\pi}\right)^{1/2} +
    (2\eta_1+d\eta_2) \mu_k \right] E_{ji}\,,
  \label{eq:sigmasymc}\\
  \sigma_{[ji]}^{(\rm c)}&=
  2^{d}g_0\nu\rho a \left(\frac{T_{\rm t}}{\pi}\right)^{1/2}\eta_2R_{ji}\,,
  \label{eq:sigmaantisymc}
\end{align}
where
\begin{align}
  \mu_k&=\frac{2(\pi T_{\rm t})^{1/2}\rho a}{\nu g_0}%
  \frac{\{
    2^{-d-1}(d+2)
    +\frac{\nu g_0}{4}
    [6\eta_1^2 - 4\eta_1 + 2 d \eta_1\eta_2 - (d-2)\eta_2 -2\eta_2^2]
    -\frac{\eta_2^2 T_{\rm r}}{2 K T_{\rm t}} \nu g_0
    \}}
  {\{
    2(3+d)\eta_1 + (d-1)(d+3)\eta_2 -6\eta_1^2 -4 d\eta_1\eta_2
    -(d^2-3)\eta_2^2 + \frac{\eta_2^2 T_{\rm r}}{K T_{\rm t}}
    \}}\,,
\end{align}
and $\nu$ is the volume fraction which may be related to
the number density $n$ of the particle as
\be
\nu=\frac{\pi^{d/2}}{\Gamma\left(\frac{d+2}{2}\right)}
a ^d n.
\ee
Now, the constitutive equation for the total stress tensor $\sigma$ is
given by
\be
\sigma_{ji}(E_{ji},R_{ji},\Omega_{kji}=0)=-\rho T_{\rm t}\delta_{ji}+\sigma_{\{ji\}}^{(\rm k)}
+\sigma_{\{ji\}}^{(\rm c)}+\sigma_{[ji]}^{(\rm c)}
\ee
with (\ref{eq:sigmasymk})--(\ref{eq:sigmaantisymc}).
From symmetry, we have
$\mu_{kji}(E_{ji},R_{ji},\Omega_{kji}=0)=0$~\footnote{Expand $\mu_{kji}(E_{ji},R_{ji},\Omega_{kji}=0)$
  in the power series of $E_{ji}$ and $R_{ji}$.
  Note that the coefficient tensors are of odd degree.  Since there is
  no special direction in the granular system, the coefficient tensor
  must be isotropic.  Isotropic tensors of odd degree are $0$.}.
The obtained constitutive equations are a restricted version of
those in~\cite{Lun1991} in the sense that constants $\pd_j v_i$,
$\omega_{ji}$, $T_{\rm t}$ and $T_{\rm r}$ are assumed, and they
are a generalized version in the sense that $d$ is arbitrary
whereas $d=3$ in~\cite{Lun1991}.

\section{Kanatani's theory}
\label{sec:Kanatani}

A micropolar fluid theory for relatively dense granular flows was
developed by Kanatani~\cite{Kanatani1979int}. The theory can be
outlined as follows. As in section~\ref{sec:kinetic}, a system of
the particles with the same radius $a$, mass $m$ and moment of
inertia $I$ is considered. Let the two contacting spherical
particles be labeled as 1 and 2. The unit vector $\bm k$ is the
same as (\ref{eq:defk}) (see figure~\ref{fig:2_grains}). From
(\ref{eq:defCW}), the quantities associated with particles,
$c_i^{(\alpha)}$ and $w_{ji}^{(\alpha)} (\alpha=1,2)$, can be
related to the macroscopic fields as \be c_{i}^{(\alpha)} =
v_{i}(\bm r^{(\alpha)}) + C_{i}^{(\alpha)}, \qquad
w_{ji}^{(\alpha)} = \omega_{ji}(\bm r^{(\alpha)}) +
W_{ji}^{(\alpha)}, \label{eq:v+v'} \ee where $C_{i}^{(\alpha)}$
and $W_{ji}^{(\alpha)}$ denote the fluctuations around the
macroscopic fields. Since $|\bm r^{(1)} -\bm r^{(2)}|=2a$ is small
compared to the typical scale in which the amplitudes of fields
vary, $v_{i}(\bm r^{(2)})$ and $\omega_{ji}(\bm r^{(2)})$ can be
approximated by its Taylor series about $\bm x^{(1)}$ up to the
first-order terms, i.e., \be v_{i}({\bm r}^{(2)}) = v_{i}+ 2 a
k_{j} \pd_j v_i\,, \qquad \omega_{ji}({\bm r}^{(2)}) = \omega_{ji}
+ 2 a k_{l} \Omega_{lji}\,,
\label{eq:vwbeta}
\ee
where we have omitted writing the argument $\bm x^{(1)}$ in
$v_i(\bm x^{(1)}), \omega_{ji}(\bm x^{(1)}),
\pd_j v_i(\bm x^{(1)})$ and $\Omega_{kji}(\bm x^{(1)})$.
From (\ref{eq:xi}), (\ref{eq:v+v'}) and (\ref{eq:vwbeta}),
the velocity difference $\xi_i$ between the two surfaces at
the contacting point is now written in terms of $\pd_j v_i$, $\omega_{ji}$,
$\Omega_{kji}$, $k_i$, $C_i^{(\alpha)}$ and
$W_{ji}^{(\alpha)}$.  Assume that the fluctuations
$C_i^{(\alpha)}$ and $W_{ji}^{(\alpha)}$ can be neglected and that
$k_i$ is isotropically distributed.  Then, we have
\be
\xi:=\langle\langle \xi_i\xi_i\rangle\rangle^{1/2}=
\left(\frac{2^3}{d}\right)^{1/2}
a \hat\omega,
\label{eq:xi_ave}
\ee
where
\begin{align}
  \hat\omega:=&\left[ \frac{d}{2(d+2)}E_{ji}E_{ji}+\frac{1}{2}R_{ji}R_{ji}
  +\frac{a^2}{2(d+2)}(\Omega_{jji}\Omega_{lli}+\Omega_{kji}\Omega_{kji}
    +\Omega_{kji}\Omega_{jki})\right]^{1/2},
\end{align}
and $\langle\langle \cdots \rangle\rangle$ denote the average over $k_i$.
Here, we have used
\begin{gather}
  \langle\langle k_{i} \rangle\rangle = 0, \qquad
  \langle\langle k_{j} k_{i}\rangle\rangle = \dfrac{1}{d} \delta_{ji}\,,
  \qquad
  \langle\langle k_{l} k_{j} k_{i}\rangle\rangle = 0 ,
  \label{eq:ntensor1}\\
  \langle\langle k_{l} k_{m} k_{j} k_{i} \rangle\rangle =
  \dfrac{1}{d(d+2)}
  (\delta_{lm} \delta_{ji} + \delta_{lj} \delta_{mi} + \delta_{li} \delta_{mj}),
  \label{eq:ntensor2}
\end{gather}
which can be derived using the general expressions of isotropic
tensors and $k_ik_i=1$. Suppose that the average energy
dissipation per unit time at a contacting point is estimated by
$\mu f \xi$, where $\mu$ is the kinetic friction coefficient and
$f$ is the average amplitude of the force applied in the direction
$\bm k$ at the contacting point. Let $N_{\mathrm{c}}$ be a number
of contacting points per a particle. Then, the energy dissipation
per unit volume per unit time $-q$ is given by \be
-q=\frac{N_{\mathrm{c}}}{2}\frac{\rho}{m}\mu f\xi,
\label{eq:q_mufxi} \ee where the factor $1/2$ is introduced to
cancel the double counting of the contacting points. Note that the
pressure $p_{\mathrm{c}}$ due to the contacting force can be
estimated by \be p_{\mathrm{c}}=\frac{\nu N_{\mathrm{c}} f}{S}\,,
\label{eq:pc} \ee where $\nu$ is the volume fraction and $S$ is
the surface area of the particle. From (\ref{eq:xi_ave}),
(\ref{eq:q_mufxi}) and (\ref{eq:pc}), one arrives at \be -q= (2
d)^{1/2}\mu p_{\mathrm{c}}\hat\omega. \label{eq:qkanatani} \ee The
estimate of $p_{\mathrm{c}}$ is given as a homogeneous function of
$E_{ji}$, $R_{ji}$ and $\Omega_{kji}$, i.e., \linebreak
$p_{\mathrm{c}}(a E_{ji},a R_{ji},a \Omega_{kji})= a^{\zeta'}
p_{\mathrm{c}}(E_{ji},R_{ji},\Omega_{kji})$ with a constant
$\zeta'$. The form of the function depends on whether the flow is
slow or fast, and here we omit the review. See the original
paper~\cite{Kanatani1979int} for the details.

Let us restrict ourselves to macroscopically uniform and steady states.
From (\ref{eq:phiq}) and (\ref{eq:qkanatani}), one obtains
\be
\Phi= (2 d)^{1/2}\mu p_{\mathrm{c}}\hat\omega,
\label{eq:PhiKanatani}
\ee
which is a homogeneous function of $E_{ji}$, $R_{ji}$ and $\Omega_{kji}$,
i.e., $\Phi(a E_{ji},a R_{ji},a \Omega_{kji})=
a^\zeta\Phi(E_{ji},R_{ji},\Omega_{kji})$ with $\zeta = \zeta'+1$.
With the help of
Euler's homogeneous function theorem,
\be
\Phi=\frac{1}{\zeta}\left(\frac{\pd \Phi}{\pd E_{ji}} E_{ji}
  +\frac{\pd \Phi}{\pd R_{ji}} R_{ji}
  +\frac{\pd \Phi}{\pd \Omega_{kji}}\Omega_{kji}\right),
\ee
the choice
\be
\sigma_{\{ji\}}=\frac{1}{\zeta}\frac{\pd \Phi}{\pd E_{ji}}\,, \qquad
\sigma_{[ji]}=\frac{1}{\zeta}\frac{\pd \Phi}{\pd R_{ji}}\,, \qquad
\frac{1}{2}\mu_{kji}=\frac{1}{\zeta}\frac{\pd \Phi}{\pd \Omega_{kji}}\,,
\label{eq:choice}
\ee
is made for constitutive equations
which are consistent with (\ref{eqn.Phi_granular}).

\section{Comments on the kinetic theory and Kanatani's theory}
\label{sec:comments}

In the kinetic theory by Lun, the binary collision is assumed,
i.e., $n$-particle collisions for $n > 2$ are neglected,
and the particle-pair distribution $f^{(2)}$ is assumed to be
approximated by a product of $g_0$ and $f^{(1)}$ as
(\ref{eq:f2gf1f1}).  These assumptions seem to be valid
for small volume fraction, $\nu \ll 1$. This is because,
in such a case,
$n$-particle collisions for $n > 2$ would be rare
and the velocities of the two particles before a binary
collision would be statistically almost independent.
Furthermore, macroscopic fields $\pd_j v_i$ and $\omega_{ji}$
are assumed to be small in comparison to their microscopic
fluctuations.  All of these assumptions would become inappropriate
with the increase of $\nu$.

Even when $\nu$ is small, there are a few points one should
consider carefully in the kinetic theory by Lun.  Regarding the
distribution function $f_0^{(1)}$ in (\ref{eq:f(1)0}), different
temperatures $T_{\rm t}$ and $T_{\rm r}$ are assigned to
transitional and rotational degrees of freedom.  This is not a
redundant setting since the violation of equipartition of energy
between transitional and rotational degrees of freedom indeed
occurs. For example, Huthmann and
Zippelius~\cite{HuthmannZippelius1997} showed that the
equipartition is immediately destroyed even if it is  initially
satisfied.
Some studies (e.g., Goldhirsh, Noskowicz and
Bar-Lev~\cite{GoldhirshNoskowiczBar-Lev2005}, Kranz et
al.~\cite{KranzBrilliantovPoeschelZippelius2009}) suggest that
there are substantial deviations from Gaussian and correlation
between $C_i$ and $W_{ji}$ in the distribution function $f^{(1)}$.
Therefore, there can be a substantial deviation from
(\ref{eq:f(1)}) with (\ref{eq:f(1)0}) and (\ref{eq:phi}) in
$f^{(1)}$.  Since our aim is to extract information for the
constitutive equations, full detailed information on $f^{(1)}$ is
not necessary. However, we do not know how much information is
sufficient for our purpose a priori. Here, it is assumed that the
effect of the corrections on moments higher than two in
$f_0^{(1)}$ is not so large.

Note that it is suggested by a number of studies that $\beta$
depends on the angle $\vartheta:=\arctan(\xi_t/\xi_n)$ of the
oblique collision of the particles, where $\xi_n= k_i\xi_i$ and
$\xi_t=\sqrt{\xi_i\xi_i-\xi_n^2}$. Walton and
Braun~\cite{WaltonBraun1986} derived
\begin{equation}
\beta\simeq
\begin{cases}
-1 +\left(1+\frac{1}{K}\right) \mu (1+e) \cot \vartheta &
(\vartheta \geqslant \vartheta_0)\\
\beta_0 &
(\vartheta < \vartheta_0)
\end{cases},
\label{eq:beta_th}
\end{equation}
where $\beta_0$ is the maximum value of the coefficient of
roughness and $\vartheta_0$ is the critical angle.  The
form of $\beta$ is verified in the numerical simulation
by Saitoh and Hayakawa~\cite{SaitohHayakawa2007}.
Therefore, the constant $\beta$ assumed in the kinetic theory
by Lun, should be regarded as an approximation.  Again,
we do not know a priori how much detailed information on
$\beta$ is required to obtain constitutive equations. It should be verified
whether the simplified model of constant $\beta$ is adequate
herein.

In Kanatani's theory, a relatively large $\nu$ is considered.
Every particle is in contact with some other particles during
almost all the time. The estimate of the energy dissipation $\mu f
\xi$ per unit time per a contact point implies that the velocity
difference $\xi$ is maintained to the order of the macroscopic
time scale.  However, it is not clear whether this is the general
case for a large $\nu$.  The velocity difference $\xi$ possibly
decays in the order of microscopic time scale during the contact
due to  friction.  Furthermore, among many choices of constitutive
equations that are consistent with a given dissipation function
$\Phi(E_{ji},R_{ji},\Omega_{kji})$, a particular choice
(\ref{eq:choice}) is made. When the order of homogeneity $\zeta$
is $2$, the choice of constitutive equations (\ref{eq:choice})
leads to the linear response satisfying Onsager's reciprocal
relations~\cite{Onsager1931_1}. Thus, the choice is valid.
However, when $\zeta\ne 2$ and the constitutive equations are
nonlinear, such a validation is not possible.

Some of the points given above will be examined by means of numerical
simulations in the next section.

\section{Numerical simulations}
\label{sec:simulation}
\subsection{Setting}
\label{sec:setting}
We performed numerical simulations of a system of granular particles
using a distinct element method (DEM).
In DEM,
the interaction forces for every contacting pair of particles are
calculated at each time step,
and the equations of the motion are solved using a difference method.
The dimension $d$ of the configuration space is $2$.
All the particles are disks of the same radius $a$.
The mass is uniformly distributed inside the particles and thus
the mass $m$ and the moment of inertia $I$ of the particles are
related as $K=I/m a^2=1/2$.
As a mechanism of the contact,
we assume Hooke's law of elasticity in the direction parallel to
$\bm k$ of (\ref{eq:defk}),
that is, $F= \kappa l$ where
$l=\max(2a -|{\bm x}^{(\beta)}-{\bm x}^{(\alpha)}|,0)$
is the overlap length and $\kappa$ is the force constant.
The kinetic friction force $\mu F$ is
applied at the contacting point in the direction to reduce
the velocity difference $\xi$
between the surfaces of the two particles (see section~\ref{sec:Kanatani}).
We have chosen the above model of contact since it is one of
the simplest models that induces both energy dissipation and
rotational degree of freedom of particles, that are essential
elements to consider the micropolar nature of the granular flow.
For simplicity, we have neglected nonlinearity in the relation between
$l$ and $F$, inelasticity in the direction parallel to $\bm k$ and
the static friction in the direction of the velocity difference $\xi$.
By virtue of  simplicity of the model,
the microscopic characteristics of  granular
particles are completely determined by four parameters,
the radius $a$, the mass $m$, the force constant $\kappa$ and
the kinetic friction coefficient~$\mu$.

$N$ particles are put in a square domain with the length of
sides $L_x$ and $L_y$.  The area fraction $\nu$ is given by
$\nu=N \pi a^2/L_xL_y$.  We fix $L_x/a=L_y/a=200$ for all the
simulations except for the case $\nu=0.1$ in which
$L_x/a=L_y/a=600$.
The time scale associated with the collision is given by
$t_{\rm col}=(m/\kappa)^{1/2}$.
The time step $\Delta t$ of DEM is set to $10^{-2} t_{\rm col}$,
which is sufficiently smaller than $t_{\rm col}$ so that
the collision process is resolved in the simulation.
Since there is no damping in the direction parallel
to $\bm k$, the time scale associated with the damping is
$t_{\rm damp}=\infty$.

For the kinetic friction coefficient, we used $\mu=0.3$ and $0.8$.
Note that  kinetic theories which eliminate the spin field as an
independent field are
suggested~\cite{JenkinsZhang2002,YoonJenkins2005} for small
kinetic friction coefficient. Therefore, the aspect of granular
flows as a micropolar fluid would become significant for large
$\mu$.  In order to investigate this aspect, we have chosen
somewhat larger $\mu$ compared to those in some measurements, e.g.
$\mu<0.2$ in~\cite{LabousRosatoDave1997}.

We employ the Lees-Edwards periodic boundary
conditions~\cite{LeesEdwards1972} for the velocity $\bm c$ of
particles,
\begin{align}
c_i(x=L,y,t)&=c_i(x=0, y,t),
\label{eq:LeesEdwards_x}\\
c_i(x,y=L,t)&=c_i\bigl( (x- L_y \dot\gamma t)\ {\rm mod}\ L_x,y=0,t\bigr)
+L_y \dot\gamma \delta_{ix}\,,
\label{eq:LeesEdwards_y}
\end{align}
where $\dot\gamma$ is a constant.
The external torque
$\tilde \tau_{ji}=\tilde \tau (\delta_{jy}\delta_{ix} -
\delta_{jx}\delta_{iy})$, constant in time,
is applied to every particle.

The macroscopic fields $\rho(x,y)$, $v_i(x,y)$ and $\omega_{ji}(x,y)$
are defined by the spatial averages,
\be
\rho(x,y)=\frac{m N_\Delta}{\Delta x\Delta y}\,, \qquad
v_i(x,y)=\frac{1}{N_\Delta}\sum_\alpha c_i^{(\alpha)},\qquad
\omega_{ji}(x,y)=\frac{1}{N_\Delta}\sum_\alpha w_{ji}^{(\alpha)},
\ee
where $\bm c^{(\alpha)}$ and $\bm w^{(\alpha)}$ are, respectively,
the velocity and the spin of the particle labeled by $\alpha$,
$N_\Delta:=\sum_{\alpha} 1$ and $\sum_{\alpha}$ denotes the summation over
the particles such that $x-\Delta x/2 \leqslant x^{(\alpha)} <x+\Delta x/2$
and $y-\Delta y/2 \leqslant y^{(\alpha)} < y+\Delta y/2$.
The notations $E_{ji}(x,y)$, $R_{ji}(x,y)$ and $\Omega_{kji}(x,y)$
are introduced similarly to (\ref{eq:EROmega}).
The simulations were performed up to the time that the velocity
field relaxes to a quasi-steady state.  For all the sets of parameters
that we investigate in this paper, the velocity field relaxed
to a uniform simple shear profile with the shear rate
$\dot\gamma$, i.e., $v_i(y)=(\dot\gamma y + A)\delta_{ix}$ where $A$
is a constant.  Here, we have chosen $\Delta x=L_x$ so that
$v_i$ is independent of $x$.  In terms of $E_{ji}(y)$, we have
\be
E_{ji}(y)=\frac{\dot\gamma}{2}
\left(\delta_{jy}\delta_{ix} + \delta_{iy}\delta_{jx}\right).
\label{eq:Dyx}
\ee
It was found in the simulations that, when $E_{ji}(y)$
relaxed to a uniform field, $\omega_{ji}(y)$ also
relaxed to a uniform field.  Consequently, we have
\be
R_{ji}(y)=R
\left(\delta_{jy}\delta_{ix} - \delta_{iy}\delta_{jx}\right),
\qquad \Omega_{kji}(y)=0.
\label{eq:Ryx}
\ee
The shear rate $\dot\gamma$
is directly controlled by the boundary conditions
(\ref{eq:LeesEdwards_x}) and (\ref{eq:LeesEdwards_y}).
We observed $R=0$,
i.e., the spin $\omega_{ji}$ is subordinate to $\pd_{[j}v_{i]}$,
when there is no external torque $\tilde \tau$.
The value of $R$ is controlled by changing the value
of $\tilde \tau$. In the present study, we varied $\tilde \tau$ in the
range $\tilde \tau \geqslant 0$.
In such cases, we observed $R\leqslant 0$, i.e., $\omega_{yx} \geqslant \pd_{[y}v_{x]}$,
at the quasi-steady states.

Let us consider a line $x_i= h$ where $h$ is a constant.
In the simulation, the kinetic contribution to the stress tensor,
$\sigma^{\rm (k)}_{ji}$, is estimated by
\be
\sigma^{\rm (k)}_{ji}(h)=-\frac{1}{L_j}
{\sum_\alpha}^{(h)} {\rm sgn}(v_j^{(\alpha)})
\frac{m v_i^{(\alpha)}}{\Delta t},
\ee
where the summation $\sum_\alpha^{(h)}$ is taken over the particles
$\alpha$ which cross the line $x_j=h$ during the time step $\Delta t$.
The contribution of contacts to the stress,
$\sigma^{\rm (c)}_{ji}$, is estimated by
\be
\sigma^{\rm (c)}_{ji}(h)=\frac{1}{L_j}
{\sum_{\alpha,\beta}}^{(h)}F_i^{\alpha \to\beta},
\ee
where $F_i^{\alpha \to \beta}$ is the contacting force applied
to the particle $\beta$ by the particle $\alpha$ and
the summation $\sum_{\alpha,\beta}^{(h)}$ is taken over the contacting pairs
$\{\alpha,\beta\}$ such that $x_j^{(\beta)}\leqslant h <x_j^{(\alpha)}$.
The mean stress $\sigma_{ji}(h)$ over a line $x_j=h$ can be averaged
over $h$ to give the mean stress $\sigma_{ji}$ of the whole domain,
\be
\sigma_{ji}=\sigma^{\rm (k)}_{ji} + \sigma^{\rm (c)}_{ji},
\ee
with
\begin{align}
\sigma^{\rm (k)}_{ji}&=-\frac{1}{L_x L_y}
\sum_{\alpha} mv_j^{(\alpha)} v_i^{(\alpha)},
\label{eq:sigmak_sim}\\
\sigma^{\rm (c)}_{ji}&=\frac{1}{L_x L_y}
\sum_{\alpha,\beta} F_i^{\alpha\to\beta}
\left( x_j^{(\alpha)} -x_j^{(\beta)}\right)
\label{eq:sigmac_sim},
\end{align}
where the summations $\sum_{\alpha}$ and $\sum_{\alpha,\beta}$
are, respectively, taken over the particles and the contacting pairs.
Note that the pairs $\{\alpha,\beta\}$ and $\{\beta,\alpha\}$ are
not distinguished and are counted only once in the
summation $\sum_{\alpha,\beta}$.

\begin{figure}
\centerline{\includegraphics[width=0.6\linewidth]{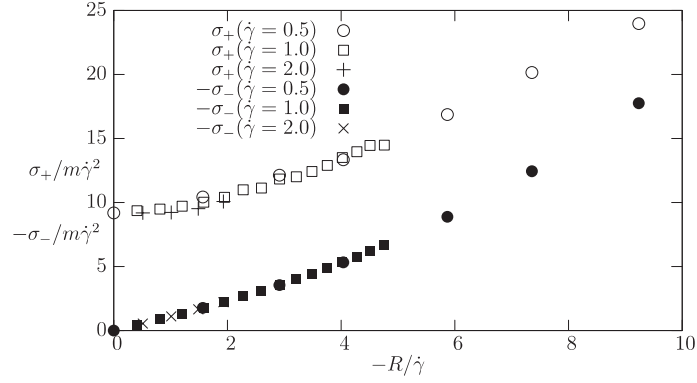}}
\caption{\label{fig:bagnold}Stresses $\sigma_{\pm}$ normalized by
$m\dot\gamma^2$ as functions of $R/\dot\gamma$ for
the mean shear $\dot\gamma t_{\rm col}=
0.5\times 10^{-3},1.0\times 10^{-3}$ and $2.0\times 10^{-3}$.
The friction coefficient $\mu$
and the area fraction $\nu$ are fixed to
$\mu=0.3$ and $\nu=0.7$.}
\end{figure}
Under the conditions (\ref{eq:Dyx}) and (\ref{eq:Ryx}), the
constitutive equations read \be
\sigma_\pm=\sigma_\pm(\dot\gamma,R), \ee where we have introduced
the notations $\sigma_{+}:=\sigma_{\{yx\}}$ and
$\sigma_{-}:=\sigma_{[yx]}$. There are three time scales involved
in the dynamics of the present system. They are
$t_{\dot\gamma}:=|\dot\gamma|^{-1}$, $t_{R}:=|R|^{-1}$ and the
time scale of collision $t_{\rm col}=(m/k)^{1/2}$. In the
simulations, we set $t_{\dot\gamma}\sim t_R \gg t_{\rm col}$. Let
us assume that, in such a case,  $\sigma_+$ and $\sigma_-$ do not
depend on $t_{\rm col}$.  Then, a dimensional consideration yields
similarity forms, \be \sigma_\pm(\dot\gamma,R) ={\rm
sgn}(\dot\gamma) m
|\dot\gamma|^2\tilde\sigma_\pm\left(\frac{R}{\dot\gamma}\right) =m
|\dot \gamma|\dot \gamma
\tilde\sigma_\pm\left(\frac{R}{\dot\gamma}\right),
\label{eq:sigma_scaling} \ee where $\tilde\sigma_\pm$ are
dimensionless functions, which may depend on the kinetic friction
coefficient $\mu$ and the area fraction $\nu$. The origin of the
function $\rm sgn(\dot\gamma)$ is a requirement of the symmetry
$\sigma_{\pm}(-\dot\gamma,-R)=-\sigma_{\pm}(\dot\gamma,R)$. Note
that, in the case of $R=0$, (\ref{eq:sigma_scaling}) reduces to
the well known Bagnold scaling~\cite{Bagnold1954},
$\sigma_+\propto \dot\gamma^2$. In figure~\ref{fig:bagnold},
$\sigma_\pm/m\dot\gamma^2$ from the simulations are plotted as a
function of $R/\dot\gamma$ for $\dot\gamma t_{\rm col}=0.5\times
10^{-3},1.0\times 10^{-3}$ and $2.0\times 10^{-3}$ ($\mu$ and
$\nu$ are fixed). The collapse of the data implies that the
similarity forms (\ref{eq:sigma_scaling}) are valid within the
parameter range of the simulations. By virtue of
(\ref{eq:sigma_scaling}), we can fix $\dot\gamma$ and vary only
$R$ to obtain the constitutive equations.  In the following
simulations, we put $\dot\gamma t_{\rm col}=1.0 \times 10^{-3}$.

\begin{figure}
\centerline{
\includegraphics[width=0.49\textwidth]{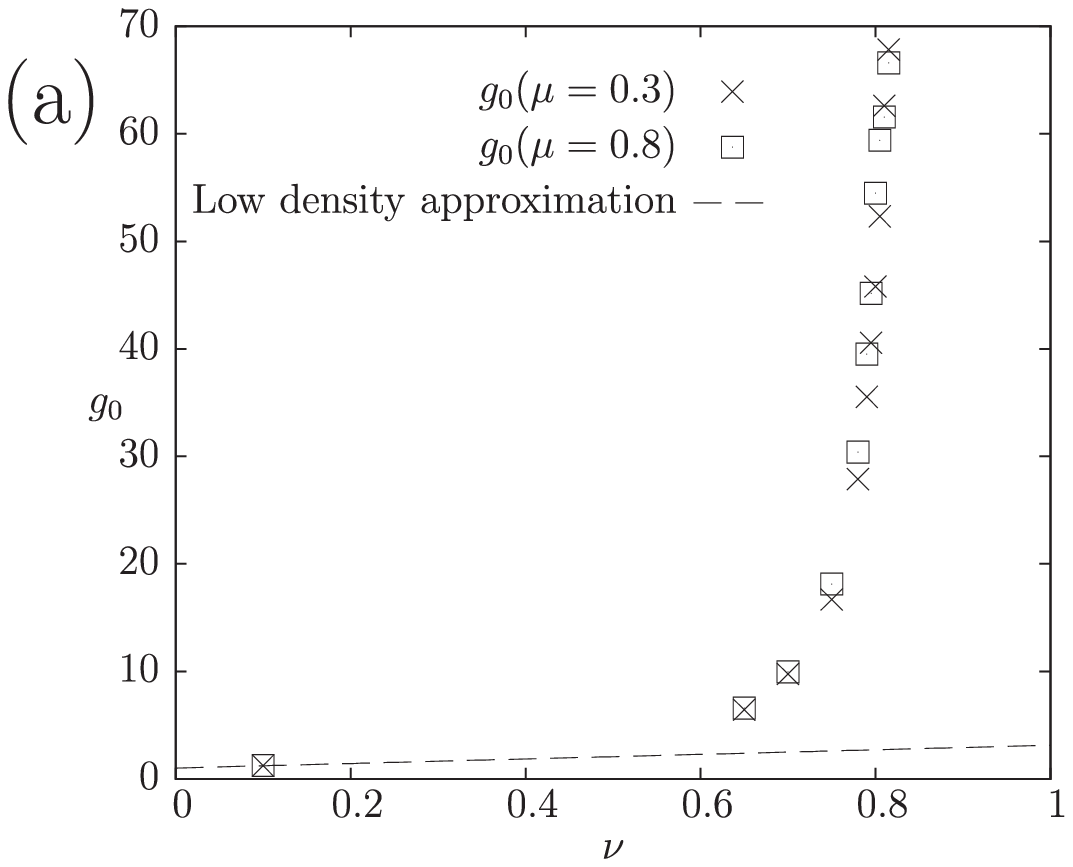}
\includegraphics[width=0.49\textwidth]{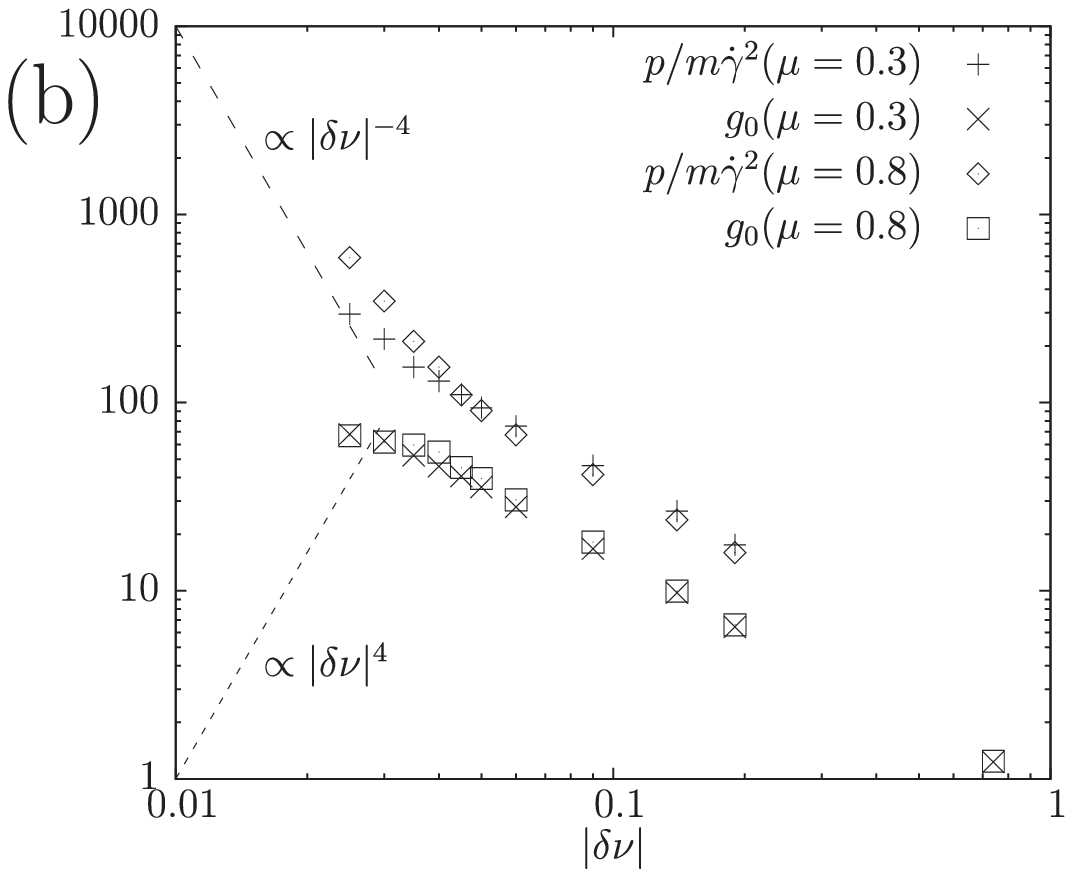}}
\caption{\label{fig:g0} (a) The value of the radial distribution
function at $r=2 a$, $g_0=g(2 a)$, in the simulations for various
$\nu$ and $R=0$.  Dashed line indicates the value of the low
density expansion to the first order. (b) The normalized pressure
$p/m\dot\gamma^2$ and the $g_0$ as functions of $|\delta
\nu|=|\nu-\nu_{\mathrm{J}}|$, where $\nu_{\mathrm{J}}=0.84$.  The
slope of the lines indicate the scalings in theory
in~\cite{OtsukiHayakawa2009PRE}.
}
\end{figure}
The value of the radial distribution function at $r=2a$,
$g_0=g_0(2a)$, in the simulations is given in
figure~\ref{fig:g0}(a).  The data are given for $R=0$, but they
are almost independent of $R$. Note that, for the elastic system
($e=1$) without friction ($\beta=-1$) and shearing
($\dot\gamma=0$), the low density expansion of $g_0$ at a thermal
equilibrium state can be given up to the first order of $\nu$ as $
g_0=1+\nu[ (8/3) -(\sqrt{3}/\pi)] $ (see, for
example,~\cite{HirschfelderCurtisBird1954}), and it is also shown
in a dashed line in figure~\ref{fig:g0} for comparison. One can
see from figure~\ref{fig:g0} that $g_0$ of the simulation is
approximated by the low density expansion for relatively low area
fraction $\nu=0.1$, but it largely deviates from the low density
expansion when $\nu \geqslant 0.6$. It is known that, when $\nu$
exceeds a critical value $\nu_{\mathrm{J}}$, which is called a
jamming point,
the system enters the jammed phase, in which the shear stress
remains nonzero in the limit of zero strain, i.e., yields stress.
When $\nu$ approaches $\nu_{\mathrm{J}}$, quantities such as
the pressure $p$ and %
$g_0$ are expected to obey certain scaling laws. It is found that
the present system suffers the phase separation between
crystallized region and fluid region for $\nu>0.815$, before
entering the jammed phase. Since it is impossible to estimate
$\nu_{\mathrm{J}}$ in the present system, we borrowed the value
$\nu_{\mathrm{J}}=0.84$, which is a round-off value of
$\nu_{\mathrm{J}}$ reported in~\cite{OtsukiHayakawa2009} for
two-dimensional poly-disperse frictionless granular system, as a
reference. In figure~\ref{fig:g0}(b), $p$ and $g_0$ are plotted as
functions of $|\delta \nu|=|\nu-\nu_{\mathrm{J}}|$ for various
$\nu$ and $R=0$. Since the crystallized phase is out of the scope
of the present paper, data for $\nu>0.815$ are omitted in the
figure. The scaling laws $p\sim |\delta \nu| ^{-4}$ and $g_0\sim
|\delta \nu|^4$ predicted by Otsuki and
Hayakawa~\cite{OtsukiHayakawa2009} for the particles with
frictionless, linear spring interactions, are indicated by lines
in the figures as a reference. Since there are linear spring and
frictional interactions in the present model, the scaling law
could be modified. However, since the present system experiences
crystallization, the scaling range is very narrow, in case it
exists.

\begin{figure}
\centerline{\includegraphics[width=0.6\linewidth]{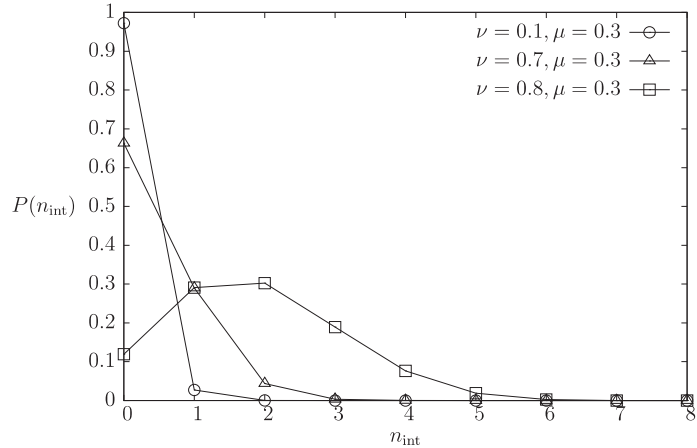}}
%
\caption{\label{fig:nbody} Probability function $P(n_{\rm int})$ of
the interfering number $n_{\rm int}$, the maximum number of particles
contacting with the colliding particle-pair simultaneously.  }
\end{figure}
For each particle-pair collision, we can define the interfering
number $n_{\rm int}$, which is the maximum number of other
particles contacting with the particle-pair simultaneously during
the collisions of the particle-pair.  $n_{\rm int}=0$ implies that
the particle-pair collision is not interfered by the other
particles.  In figure~\ref{fig:nbody}, the probability $P(n_{\rm
int})$ is given for $\nu=0.1,0.7$ and $0.8$ in the case of
$\mu=0.3$.  The data for $\mu=0.8$ are omitted since they almost
collapse with those of $\mu=0.3$.  The fact that $P(n_{\rm int})$
has a peak at $n_{\rm int}=0$ for $\nu=0.1$ and $0.7$ is
consistent with the kinetic theory. However, the peak shifts to
$n_{\rm int}=2$ for $\nu=0.8$. This suggests the failure of the
applicability of the kinetic theory based on binary collisions.
Note that the coordination number $Z$ at $\nu=0.8$ is about $0.6 -
0.7$ and it is still much smaller than the value $Z=d+1=3$ for the
isostatic condition~\cite{Edwards1998}.

Summarizing the above observations,
we can consider that $0.7 \leqslant \nu \leqslant 0.8$ is located in
an intermediate regime where $\nu$ is neither low enough
so that the kinetic theory is applicable nor
it is high enough so that the system is quite near
the jamming point $\nu_{\mathrm{J}}$.

The above conclusion might seem to be inconsistent with the
results of the numerical simulations of frictional granular shear
flows by Otsuki and Hayakawa~\cite{OtsukiHayakawa2010}. According
to~\cite{OtsukiHayakawa2010}, when friction is introduced to the
system, the critical point $\nu_J$ of the area fraction splits
into two points $\nu_{\mathrm{S}}$ and $\nu_{\mathrm{L}}$, where
the former belongs to the solid branch and the latter liquid
branch. The point $\nu=0.80$ for $\mu=0.8$ is located in the
region $\nu_{\mathrm{S}}< \nu <\nu_{\mathrm{L}}$, which implies
that the selection of the solid or liquid phase depends on
$\dot\gamma t_{\rm col}$. Although the figures for $\dot\gamma
t_{\rm col}=10^{-3}$, which is the value in the present study, are
not given in~\cite{OtsukiHayakawa2010}, it can be located in the
solid (jammed) phase.  Such a discrepancy between the results of
the present simulations and those in~\cite{OtsukiHayakawa2010} can
occur since the details of the system are different in the present
study and~\cite{OtsukiHayakawa2010}. For example, the radius of
particles is unique in the present study whereas particles with
four different radii are present in~\cite{OtsukiHayakawa2010}; the
normal viscosity is not introduced in the present study whereas
the normal viscous constant is presumably non-zero (although the
actual value is not given) in~\cite{OtsukiHayakawa2010}; and
contacting particles are always slipping in the tangential
direction in the present study whereas contacting particles can
stick when the normal contact force is strong enough
in~\cite{OtsukiHayakawa2010}. Especially, the latter fact that the
particles are always slipping in the present study would act to
raise the value of $\nu_J$ ($\nu_{\mathrm{S}}$ or
$\nu_{\mathrm{L}}$) from those in~\cite{OtsukiHayakawa2010}.

\subsection{Comparison with the kinetic theory}
\label{sec:comparisonkinetictheory}

\begin{figure}[ht]
\centerline{\includegraphics[width=0.49\textwidth]{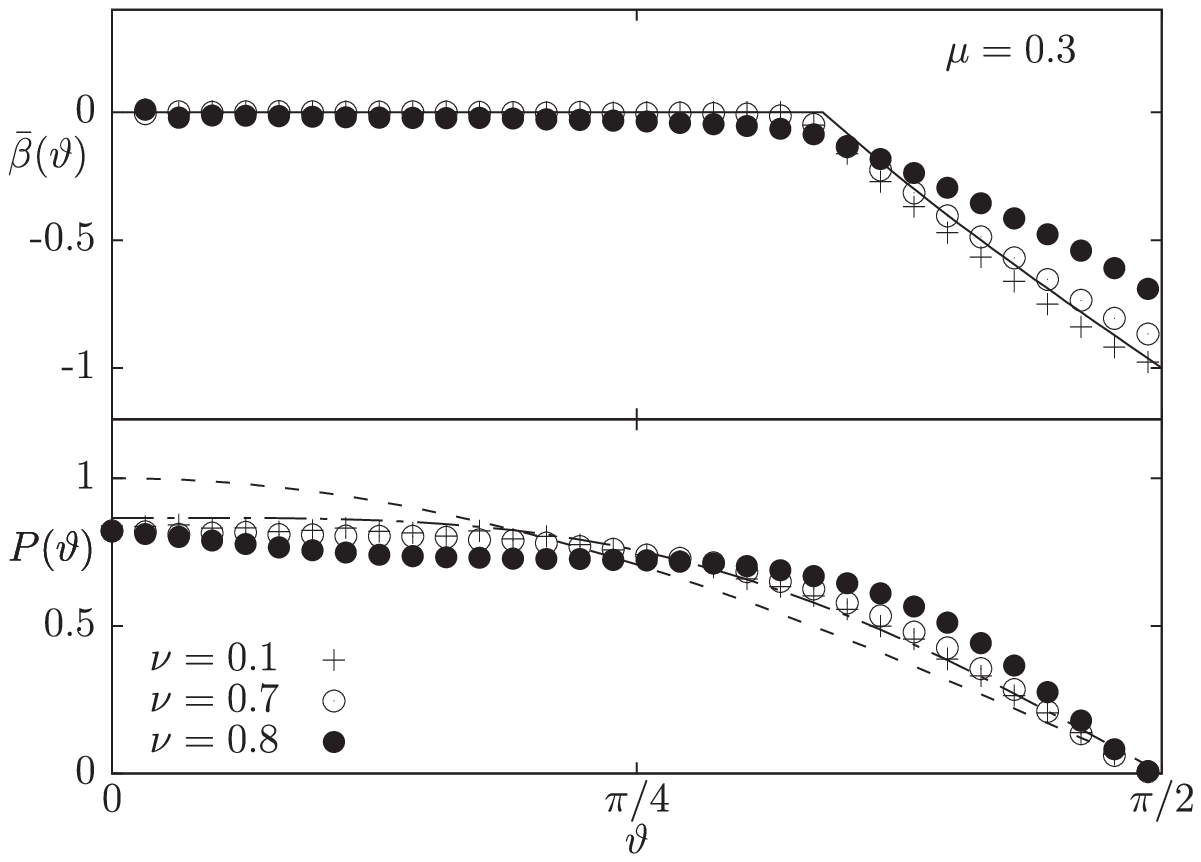}
\includegraphics[width=0.49\textwidth]{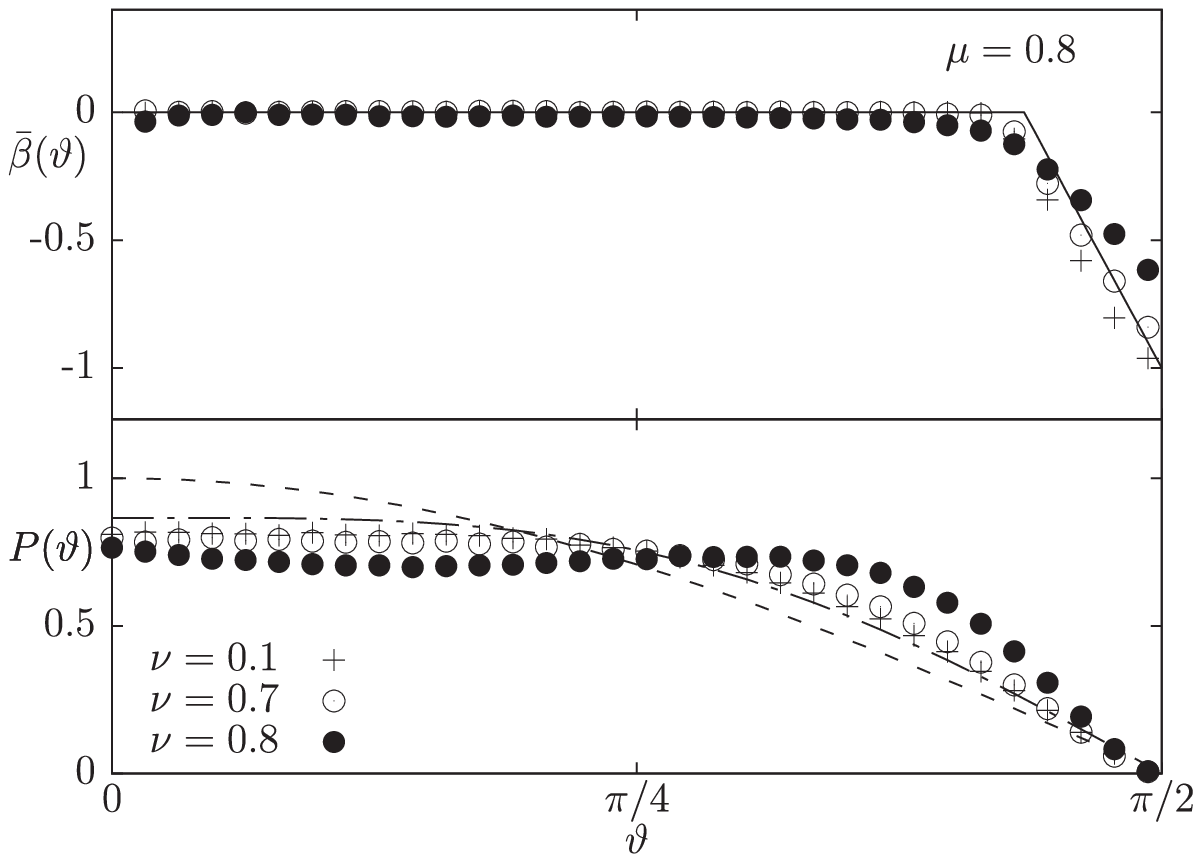}
}
\caption{\label{fig:angle}
Averaged coefficient of roughness $\bar\beta(\vartheta)$ as a function
of the oblique collision angle $\vartheta$ for various $\nu$ and $\mu$.
The solid line indicates the theoretical function (\ref{eq:beta_th}).
The dashed line shows $P(\vartheta)=\cos\vartheta$, the form for
the random collisions without spin. The dot-dashed line shows
$P(\vartheta)$ of the modeled collision with $\varsigma=0.5$.
}
\end{figure}
\begin{figure}
\centerline{\includegraphics[width=0.49\linewidth]{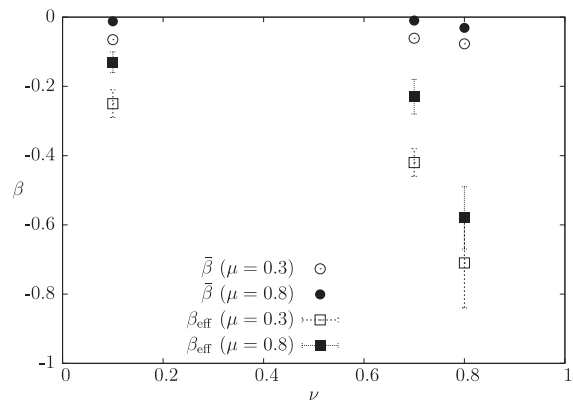}}
\caption{
\label{fig:beta}
Averaged coefficient of roughness $\bar\beta$ obtained from
the simulations (circle symbols) and
the coefficient of roughness $\beta_{\rm eff}$ fitted to
the simulation data on the basis of the kinetic theory by Lun
(square symbols) for various $\nu$ and $\mu$.
The error bars indicate $\Delta \beta$.
}
\end{figure}

In order to compare the simulation results with the kinetic theory
by Lun, we need to determine the coefficient of roughness $\beta$
in the simulations. In the simulations, we define $\beta$ by
\begin{align}
(\delta_{ij}-\check{k}_i \check{k}_j) \check{\xi}_j&=
-\beta (\delta_{ij} - k_i k_j) \xi_j\,,
\label{eq:beta_def2}
\end{align}
where $\check{\bm k}$ is the unit vector parallel to $\check{\bm
r}^{(2)}-\check{\bm r}^{(1)}$.  Since a collision has a finite
duration in the simulations, $\bm k$ and $\check{\bm k}$ are
distinguished here. The particle-pair possibly contacts with other
particles during their own collision, especially when the area
fraction $\nu$ is high and the duration of the collision is long.
$\bar\beta (\vartheta)$ is defined as the average of $\beta$ over
the particle pairs with the oblique collision angle $\vartheta$.
In figure~\ref{fig:angle}, $\bar\beta(\vartheta)$ is plotted with
the probability density function $P(\vartheta)$ of $\vartheta$.
The data is for the simulation without the external torque and the
dependency of $\bar\beta(\vartheta)$ on the external torque is
weak (the figure omitted).  For all $(\nu,\mu)$ we investigated,
we have $-1\leqslant \bar\beta(\vartheta) \leqslant 1$ which is
consistent with the assumption $-1 \leqslant \beta \leqslant 1 $
in the kinetic theory. When $\nu=0.1$, $\bar\beta$ is in agreement
with the theoretical estimate (\ref{eq:beta_th}) with $\beta_0=0$.
As $\nu$ increases, $\bar\beta(\vartheta)$ deviates from the
estimate (\ref{eq:beta_th}).  The deviation is in the direction of
decreasing the $\vartheta$ dependence of $\bar\beta(\vartheta)$.
This deviation can be understood as the effect of the collision of
the particle-pair with the other particles during the collision
within the particle-pair. The other particle randomly changes the
velocity difference $\check{\bm \xi}$ just after the particle-pair
collision. Therefore, $\beta$ becomes a random variable for fixed
$\vartheta$. For reference, the standard deviations $\Delta \beta$
at $\vartheta=1$ are $0.004(\nu=0.1), 0.05(\nu=0.7),
0.31(\nu=0.8)$ for $\mu=0.3$ and $0.0008(\nu=0.1), 0.05(\nu=0.7),
0.14(\nu=0.8)$ for $\mu=0.8$. As expected, $\Delta \beta$
increases with $\nu$.

In figure~\ref{fig:angle},  the dashed line shows
$\cos \vartheta$, the form for the random collisions without spin.
Particles tend to collide with large angle $\vartheta$ when spin
and frictional interaction are introduced.  This can be understood by
a simple model of a particle-pair with the same spin $w$ colliding
with each other with the relative velocity $\bm c$.  Let $b$ be
the impact parameter and $\Theta (-\pi/2 \leqslant \Theta \leqslant \pi/2)$
the angle between $\bm c$ and $\bm k$.  Then, we have
\begin{equation}
b=2 a \sin \Theta, \qquad
\tan \vartheta=\tan \Theta -\frac{\varsigma}{\cos \Theta}\,,
\end{equation}
where $\varsigma:=2 a w/c$.  Assuming the uniform distribution of
$b$ in $-a\leqslant b \leqslant a$, we have for the probability density
$\tilde P(\vartheta)$ of $\vartheta$
($-\pi/2 \leqslant \vartheta \leqslant \pi/2$) as
\begin{equation}
\tilde P(\vartheta)=
\frac{1}{2}\frac{1}{(1+\tan^2\Theta)
(\sqrt{1+\tan^2\Theta}-\varsigma\tan\Theta)\cos\vartheta}\,,
\end{equation}
with
\begin{equation}
\tan\Theta=\frac{\tan\vartheta +
\varsigma\sqrt{\tan^2\vartheta + (1-\varsigma^2)}}{1-\varsigma^2}\,.
\end{equation}
In figure~\ref{fig:angle}, the symmetrized probability density function
$P(\vartheta):=\tilde P(\vartheta)+\tilde P(-\vartheta)
(0\leqslant \vartheta \leqslant \pi/2)$ with $\varsigma=0.5$ is given
in a dot-dashed line for reference.
The model probability density function roughly fits the
corresponding simulation results for $\nu=0.1$ and $0.7$.
In this simple model, $c$ and $w$ are represented by constants
and their probability density functions are not considered.
If we take $c$ and $w$ as the root mean square of the fluctuations
in the simulations, we have $\varsigma\sim 0.7$ and it is not far
from $\varsigma=0.5$.  Thus, $P(\vartheta)$ for
$\nu \leqslant 0.7$ can be roughly explained by the effect of spin.
However, it is found that $P(\vartheta)$ of $\nu=0.8$ is hard to
be fitted by that of the model.  Especially, the bump around
$\vartheta=0$ is not explained by the model.  It seems that
$n$-particles interaction with $n>2$ should be taken into account in
order to understand $P(\vartheta)$ for $\nu=0.8$.

In the kinetic theory by Lun, $\beta$ is treated as a constant.
Whether we can, or cannot, approximate $\vartheta$ dependent
$\bar\beta(\theta)$ by a constant is not obvious.
Here, we take a simple average,
$\bar\beta:=\int_0^{\pi/2} \md \vartheta \bar\beta(\vartheta) P(\vartheta)$.
The values of $\bar\beta$ for various $\nu$ and $\mu$ are given
in figure~\ref{fig:beta}.

%
%
%
%
\begin{figure}[ht]
\includegraphics[width=0.49\textwidth]{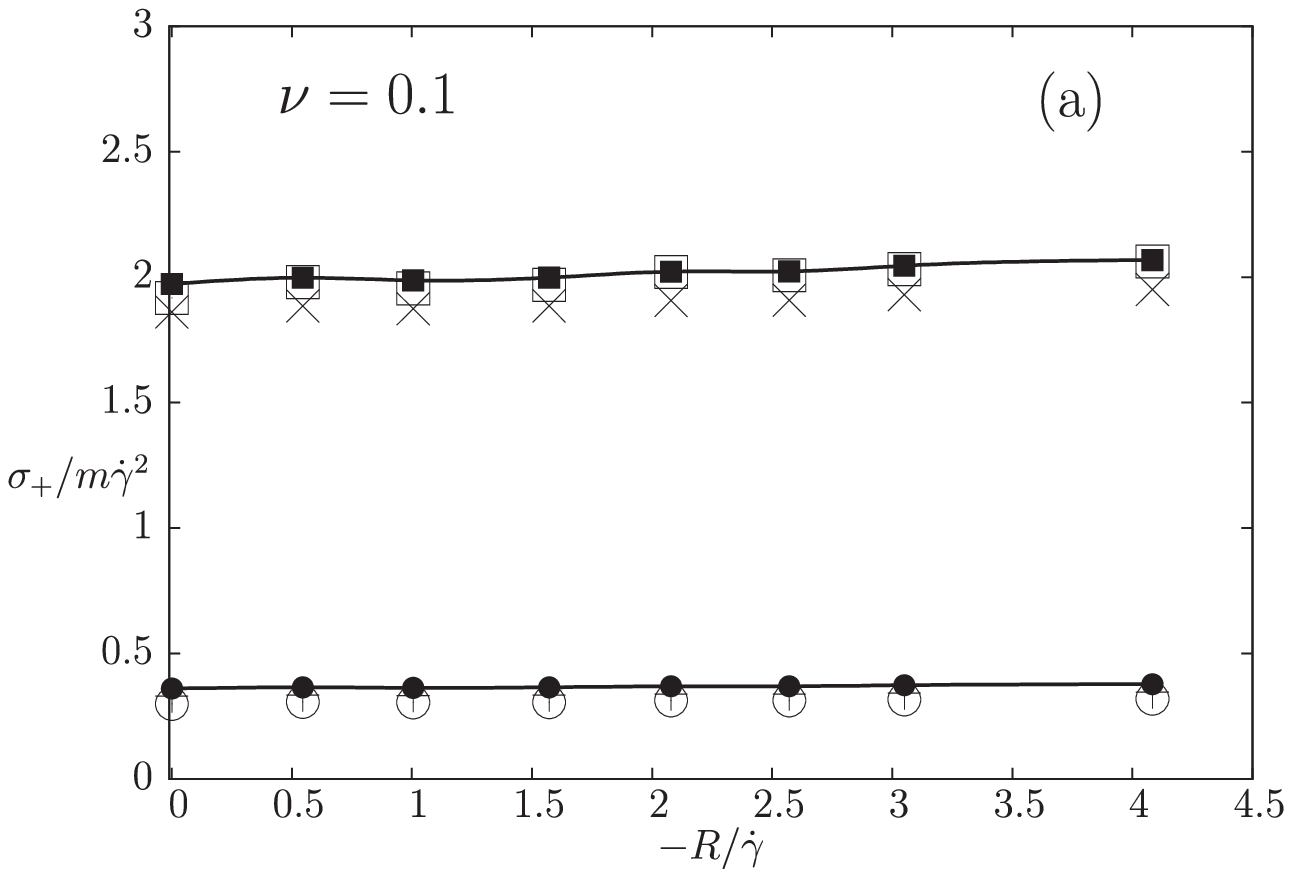}
\includegraphics[width=0.49\textwidth]{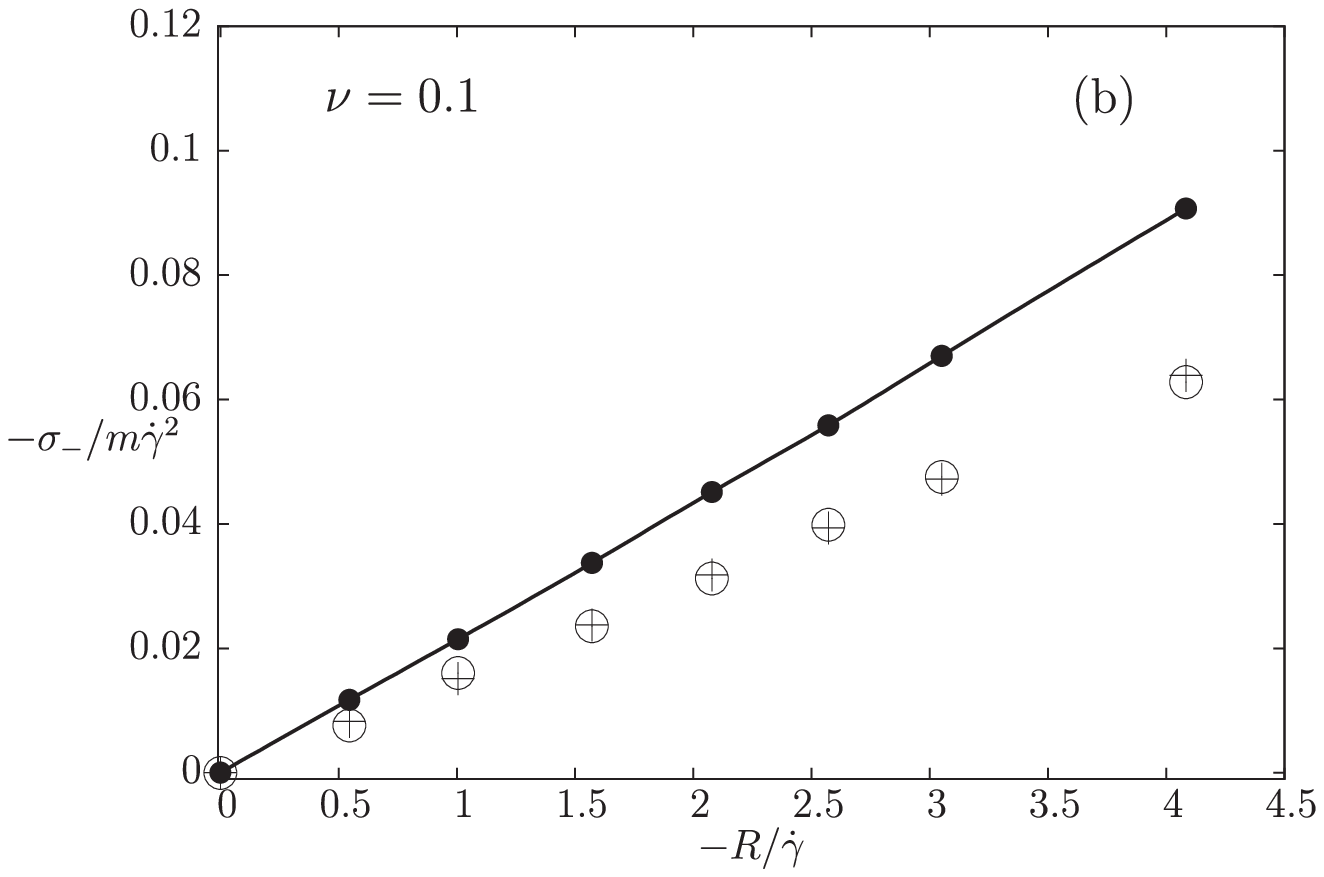}
\\
\includegraphics[width=0.49\textwidth]{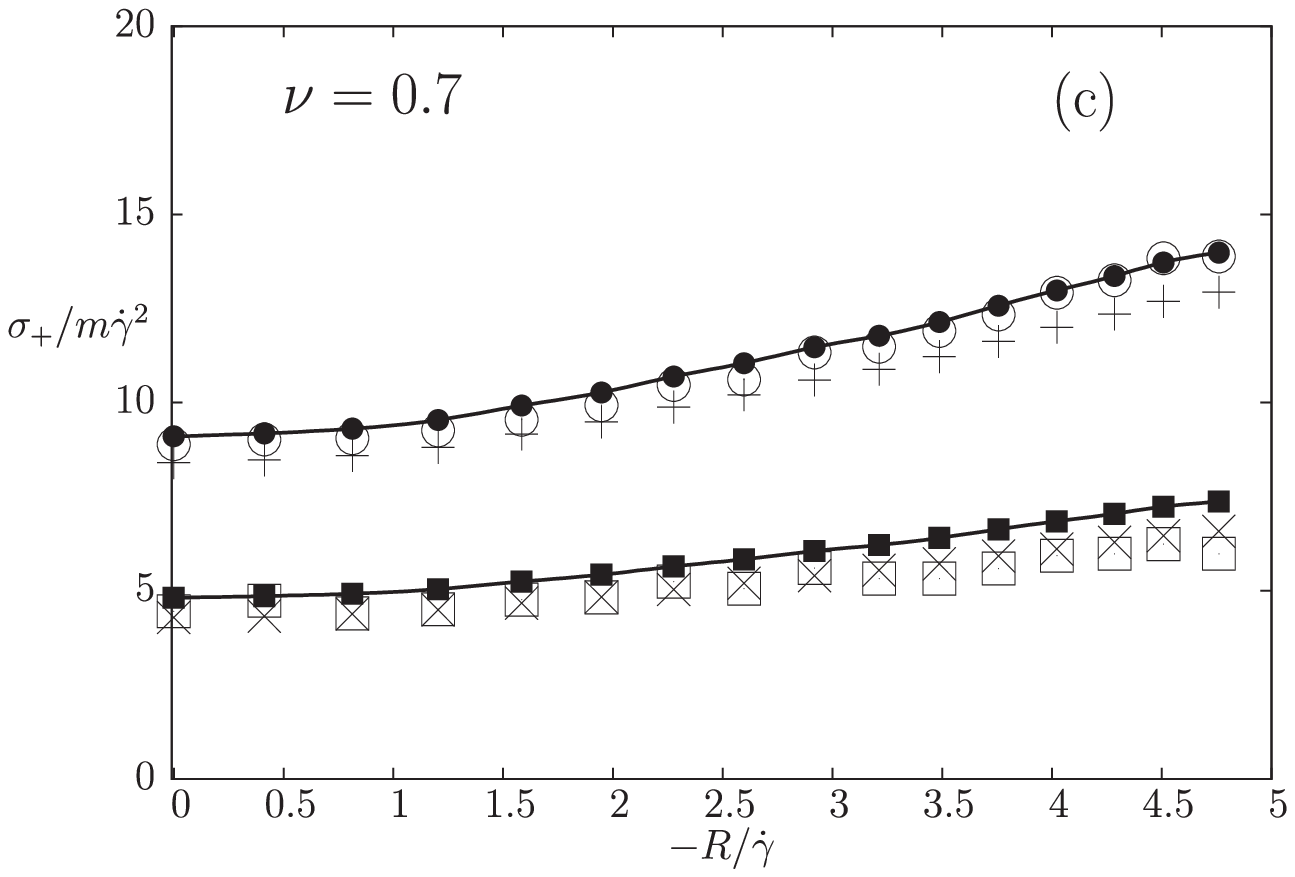}
\includegraphics[width=0.49\textwidth]{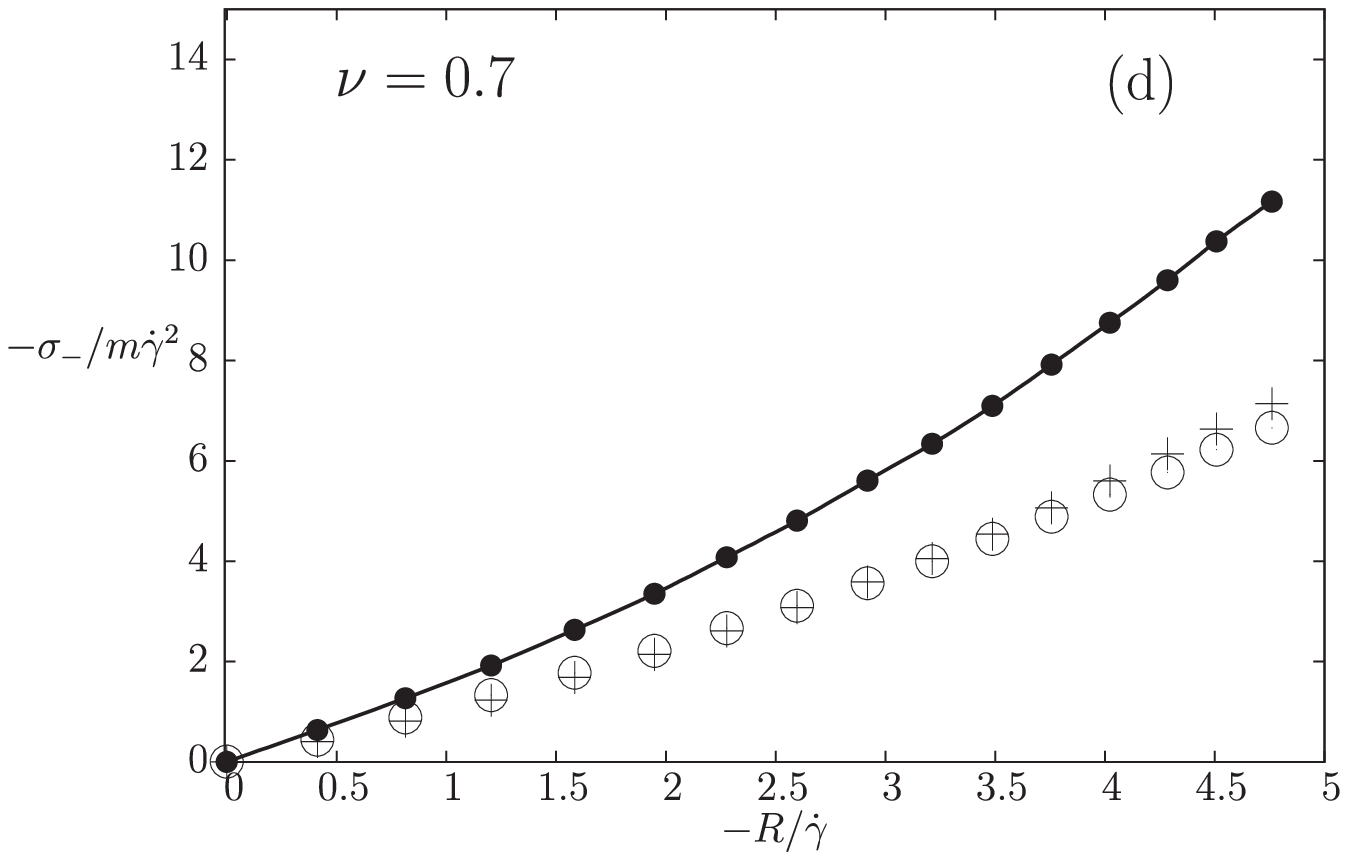}
%
\caption{\label{fig:sigmapmMU03}
Stresses $\sigma_+^{(\rm c)}$,
$\sigma_+^{(\rm k)}$ [(a),(c)], and $-\sigma_-$ [(b),(d)]
normalized by $m\dot\gamma^2$ as functions of $-R/\dot\gamma$ for
the area fractions $\nu=0.1$ and $0.7$.
The kinetic friction coefficient $\mu=0.3$ for all the figures.
Shown in (a),(c) are $\sigma_+^{(\rm c)}$ and
$\sigma_+^{(\rm k)}$ in the simulation
(white circle and square), in the kinetic theory
(\ref{eq:sigmasymk}), (\ref{eq:sigmasymc})
with $\beta=\bar\beta$
(black circle and square, interpolated with line) and
in the kinetic theory with $\beta=\beta_{\rm eff}$
(vertical and diagonal cross).
Shown in (b),(d) are $\sigma_-$ in the
simulation (white circle), in the
kinetic theory (\ref{eq:sigmaantisymc}) with
$\beta=\bar\beta$ (black circle, interpolated with line) and in
the kinetic theory with $\beta=\beta_{\rm eff}$
(vertical cross).
}
\end{figure}
%
\begin{figure}
\includegraphics[width=0.49\textwidth]{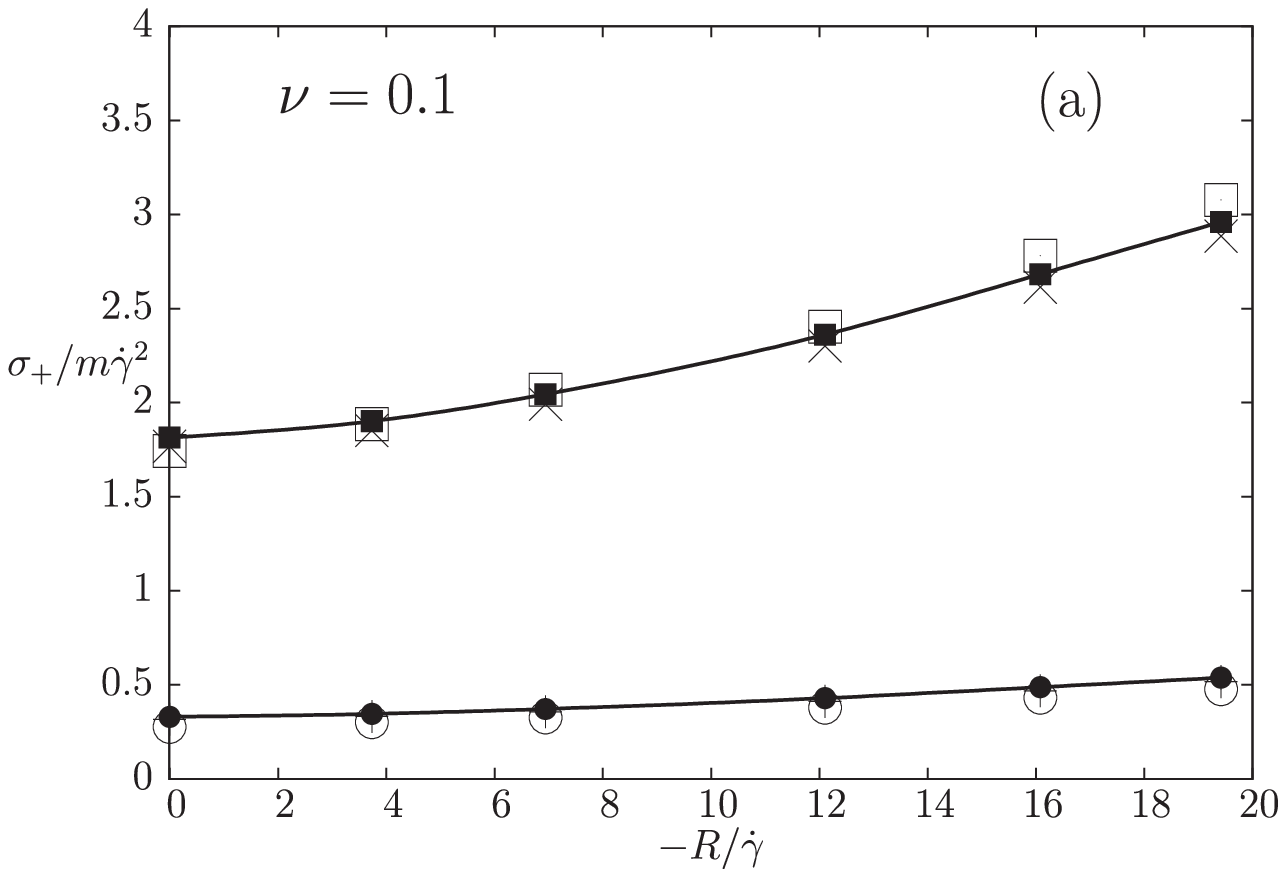}
\includegraphics[width=0.49\textwidth]{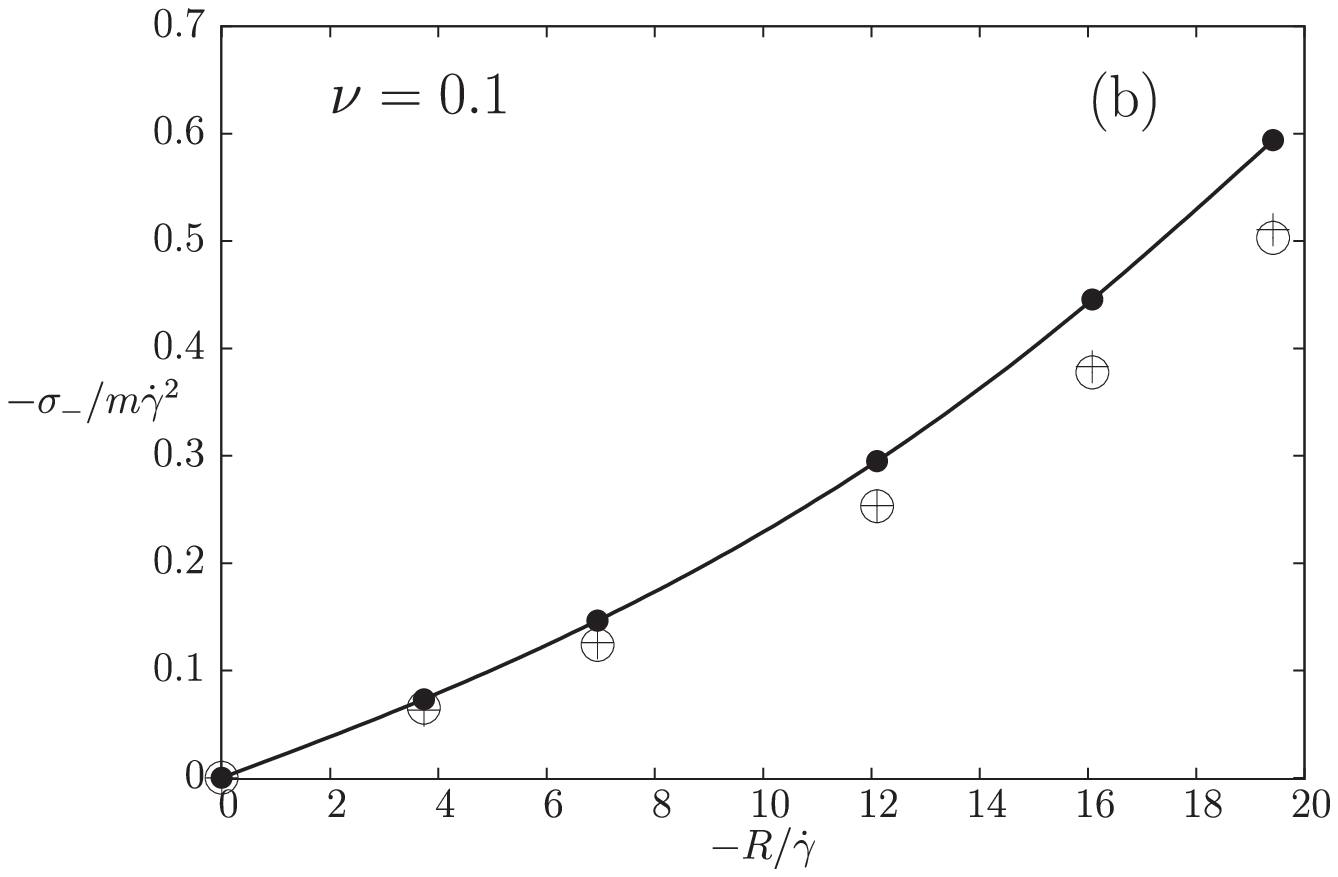}
\\
\includegraphics[width=0.49\textwidth]{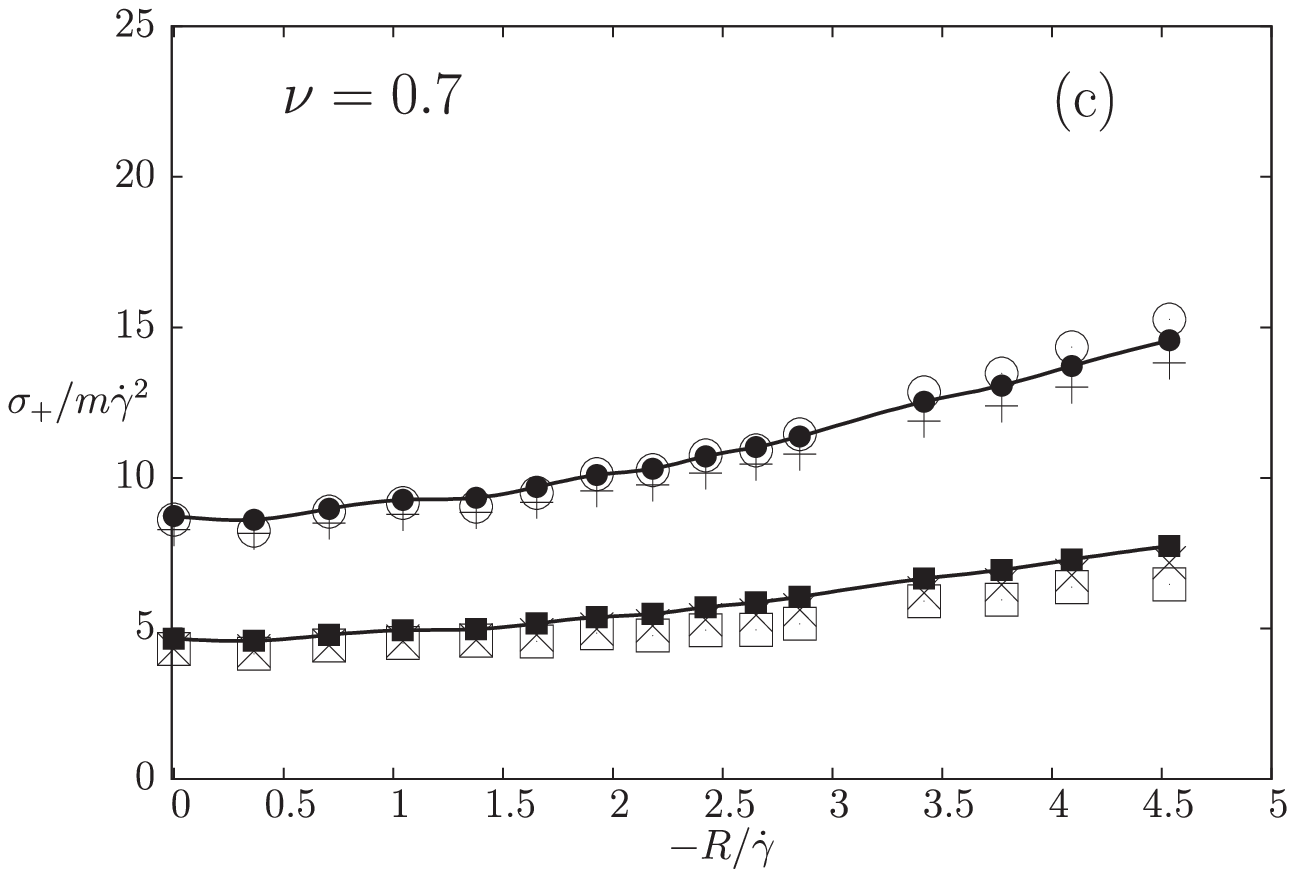}
\includegraphics[width=0.49\textwidth]{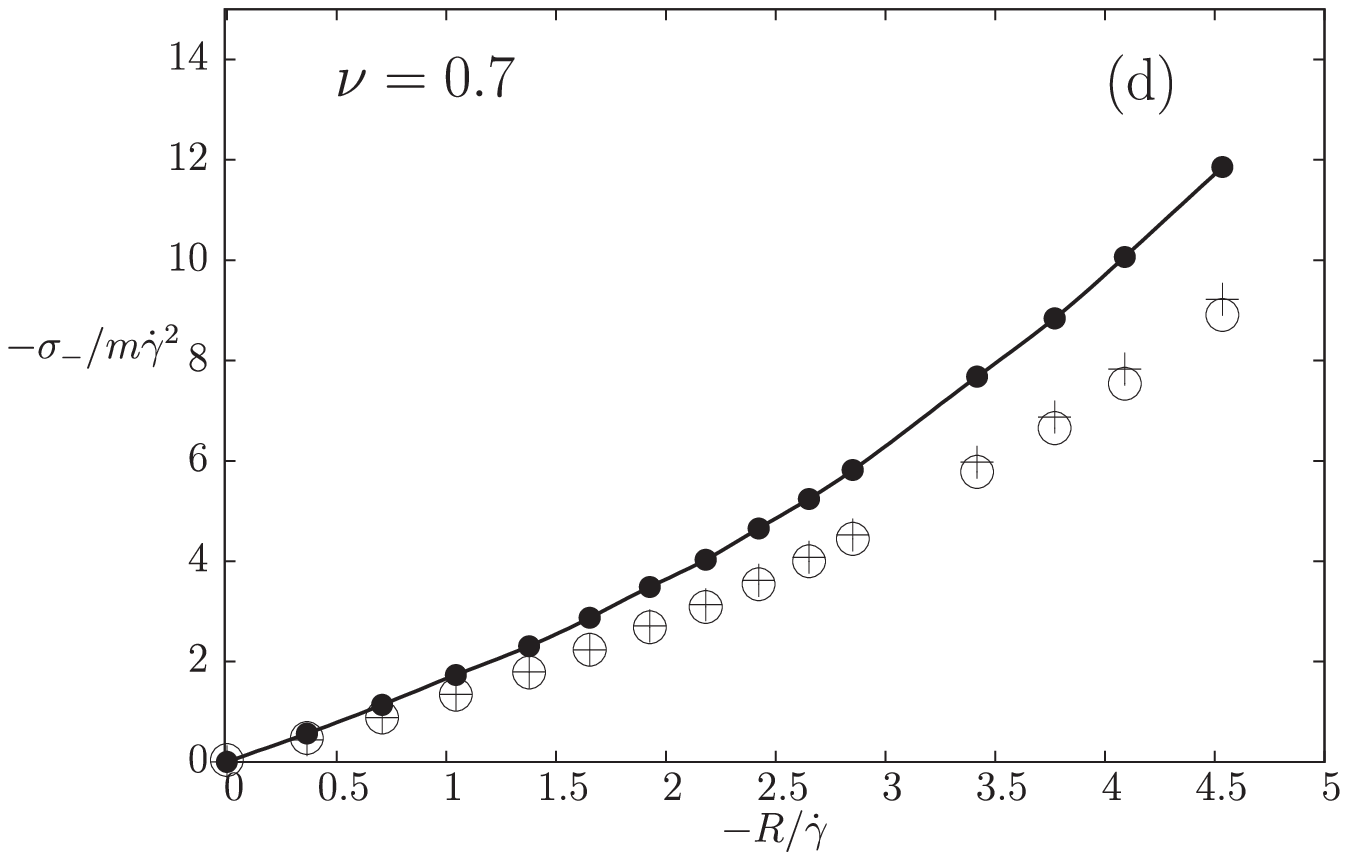}
\caption{\label{fig:sigmapmMU08}
Stresses $\sigma_+^{(\rm c)}$,
$\sigma_+^{(\rm k)}$ [(a),(c)], and $-\sigma_-$ [(b),(d)]
normalized by $m\dot\gamma^2$ as functions of $-R/\dot\gamma$ for
the area fractions $\nu=0.1$ and $0.7$.
The kinetic fraction coefficient $\mu=0.8$ for all the figures.
Shown in (a),(c) are $\sigma_+^{(\rm c)}$ and
$\sigma_+^{(\rm k)}$ in the simulation
(white circle and square), in the kinetic theory
(\ref{eq:sigmasymk}), (\ref{eq:sigmasymc})
with $\beta=\bar\beta$
(black circle and square, interpolated with line) and
in the kinetic theory with $\beta=\beta_{\rm eff}$
(vertical and diagonal cross).
Shown in (b),(d) are $\sigma_-$ in the
simulation (white circle), in the
kinetic theory (\ref{eq:sigmaantisymc}) with $\beta=\bar\beta$
(black circle, interpolated with line) and in the kinetic theory with $\beta=\beta_{\rm eff}$
(vertical cross).
}
\end{figure}
The stresses $\sigma_+^{(\rm c)}$, $\sigma_+^{(\rm k)}$, and
$\sigma_-(=\sigma_-^{(\rm c)})$ normalized by $m\dot\gamma^2$ as
functions of $R/\dot\gamma$ obtained in the simulation are given
for various $\nu(=0.1,0.7$ and $0.8)$ and $\mu(=0.3$ and $0.8)$ in
figures~\ref{fig:sigmapmMU03} and~\ref{fig:sigmapmMU08}. The
constitutive equations~(\ref{eq:sigmasymk}),~(\ref{eq:sigmasymc})
and~(\ref{eq:sigmaantisymc}) from the kinetic theory are also
given in the figures.  Here, the values obtained in the
simulations are used for the ``temperatures'' $T_{\rm t}$ and
$T_{\rm r}$, and the value of the radial distribution function at
$r=2a$, $g_0=g_0(2a)$. The coefficient of restitution $e$ is set
to $1$ following the situation of the simulation.  $\beta$ is set
to be $\bar\beta$.

When the area fraction is as small as $\nu=0.1$,
the kinetic theory is in good agreement with the simulation results
with regard to $\sigma_+^{(\rm k)}$ where the relative deviations are
within $5\%$.  The relative deviations for $\sigma_+^{(\rm c)}$ is
about $20\%$.  However, since $|\sigma_+^{(\rm k)}|$ is much larger
than $|\sigma_+^{(\rm c)}|$, the deviations for
$\sigma_+(=\sigma_+^{(\rm k)}+\sigma_+^{(\rm c)})$ are almost determined
by those of $\sigma_+^{(\rm c)}$.
The agreement between the kinetic theory and the simulation for
$\sigma_-$ is not as good as that of $\sigma_+$.  The relative
deviations are about $40\%$ for $\mu=0.3$ and $20\%$ for $\mu=0.8$.
This may be due to the approximation of the constant $\beta$
in the theory. A better agreement for $\mu=0.8$ compared to
that for $\mu=0.3$ can be explained by the fact that
the range of $\vartheta$ in which $\bar\beta(\vartheta)$
can be approximated by a constant is wider for $\mu=0.8$.

As $\nu$ increases, the ratio $\sigma_+^{(\rm c)}/\sigma_+^{(\rm k)}$
increases.  We have $\sigma_+^{(\rm c)}/\sigma_+^{(\rm k)} \sim 20$ for
$\nu=0.7$. In spite of this high ratio, the kinetic theory still gives
fairly good estimates of $\sigma_+^{(\rm k)}$ and $\sigma_+^{(\rm c)}$
whose relative deviation from the simulation results are about
$20\%$ for $\sigma_+^{(\rm k)}$ and $10\%$ for $\sigma_-^{(\rm c)}$.
As expected from the fact that the peak of $P(n_{\rm int})$ is
shifted from $n_{\rm int}=0$ for $\nu=0.8$ (see figure~\ref{fig:nbody}),
the kinetic theory is no more appropriate for $\nu=0.8$
and the formal application yields the overestimates of
$\sigma_+^{(\rm c)}$ by about $40\%$ or higher.
For $\sigma_-$, the overestimate 
is larger by the factor of more than $2$ (the figures omitted).

Now, let us find the effective $\beta_{\rm eff}$ that fits best
with the constitutive relations obtained in the simulation.
For every set of $(\mu,\nu)$, we chose $\beta=\beta_{\rm eff}$ where
$\beta_{\rm eff}$ is the value of $\beta$ which minimizes the
summation of squared relative errors, i.e.,
\begin{equation}
C(\beta)=\sum_i
\left\{
\frac{\left[\sigma_+^{\rm th}(R_i;\beta)-
\sigma_+^{\rm sim}(R_i)\right]^2}
{\sigma_+^{\rm sim}(R_i)^2}+
\frac{\left[\sigma_-^{\rm th}(R_i;\beta)-
\sigma_-^{\rm sim}(R_i)\right]^2}
{\sigma_-^{\rm sim}(R_i)^2}
\right\},
\end{equation}
with the superscripts `th' and `sim' denoting the kinetic theory
and the simulation respectively, and the subscript $i$
the index of simulations with different values of $R$.
We excluded $i$ with $R_i=0$ in the summation for $\sigma_-$ since
$\sigma_-(R=0)$ is ideally $0$ and thus it is not appropriate
to use the relative error.
In the vicinity of $\beta=\beta_{\rm eff}$, the function $C(\beta)$
can be approximated by a quadratic function of $\beta$, i.e,
$C(\beta)\simeq \tilde C(\beta)=\alpha (\beta-\beta_{\rm eff} )^2+
C_{\rm min}$.  We determined the error $\Delta \beta$ of $\beta$ to satisfy
$\tilde C(\beta\pm\Delta \beta)=2 C_{\min}$.

The obtained values of $\beta_{\rm eff}$ and $\Delta \beta$ are
given in figure~\ref{fig:beta} together with $\bar\beta$. Stresses
$\sigma_+^{(\rm k)}, \sigma_+^{(\rm c)}$ and $\sigma_-$ obtained
from~(\ref{eq:sigmasymk}),~(\ref{eq:sigmasymc})
and~(\ref{eq:sigmaantisymc}) with $\beta=\beta_{\rm eff}$ are
given in figures~\ref{fig:sigmapmMU03} and~\ref{fig:sigmapmMU08}.
One can see from the figures that $\sigma_-$ can be well fitted to
the simulation data by adjusting $\beta$.  On the other hand,
$\sigma_+$ is less sensitive to $\beta$ and so  the deviation from
the simulation data is not effectively reduced by adjusting
$\beta$. The values of $\beta_{\rm eff}$ are considerably smaller
than $\bar\beta$ for all $\nu$ and $\mu$ that we have
investigated. The discrepancy between $\bar\beta$ and $\beta_{\rm
eff}$ becomes larger as $\nu$ becomes larger.

\subsection{Comparison with Kanatani's theory}
\label{sec:comparisonkanatanitheory}
\begin{figure}
%
\centerline{\includegraphics[width=0.6\linewidth]{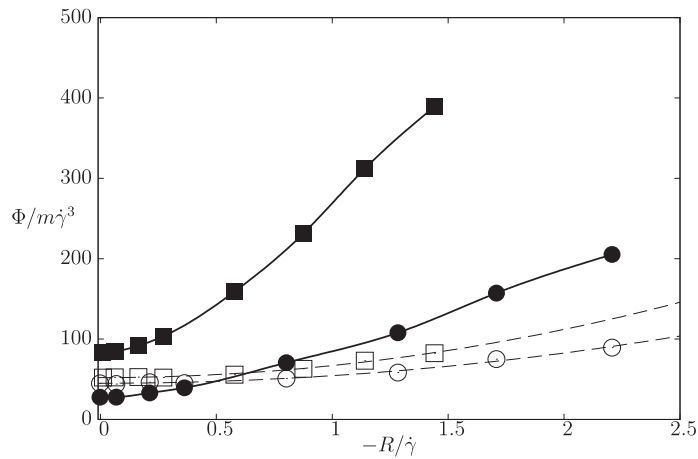}}
\caption{\label{fig:Phi} Dissipation function $\Phi$ normalized by
$m\dot\gamma^3$ as a function of $-R/\dot\gamma$ in the simulation
for $\mu=0.3$ (white circle) and $\mu=0.8$ (white square). Dashed
lines are fitted functions of the form~(\ref{eq:Phifit}).
Also shown are the normalized dissipation
function~(\ref{eq:PhiKanatani}) in Kanatani's theory with the
value of $p_{\mathrm{c}}$ from the simulation for $\mu=0.3$ (black
circle) and $\mu=0.8$ (black square). As a guide, interpolations
of the calculated data points are given in solid lines. }
\end{figure}
As we have confirmed in the previous subsection,
the estimates from the kinetic theory are in fairly good agreement
with the simulation results up to $\nu=0.7$.  When $\nu=0.8$, the
peak of $P(n_{\rm int})$ is not $n_{\rm int}=0$, that is, most pair
collisions are interfered by the other particles.  Consequently,
the kinetic theory is not applicable.  In this subsection,
we examine the applicability of Kanatani's theory for relatively
dense granular flows to the results of the simulation for $\nu=0.8$.

Under the conditions~(\ref{eq:Dyx}) and~(\ref{eq:Ryx}), the
dissipation function $\Phi$ of (\ref{eqn.Phi_granular}) reduces to
\be \Phi(\dot\gamma,R) = \sigma_+ (\dot\gamma,R) \dot\gamma + 2
\sigma_- (\dot\gamma,R) R. \label{eq:phi_gammaR} \ee In
figure~\ref{fig:Phi}, $\Phi$ obtained from the simulation is
plotted with the estimate~(\ref{eq:PhiKanatani}) of $\Phi$ in
Kanatani's theory.  In the estimate, the values from the
simulation are used for $p_{\mathrm{c}}$.  One can see from the
figure that they are not in agreement. Note that the
estimate~(\ref{eq:PhiKanatani}) depends  linearly on $\mu$ besides
the possible $\mu$ dependence in $p_{\mathrm{c}}$. However, $\Phi$
from the simulations depends on $\mu$ less sensitively. It is
suggested from the disagreement of the $\mu$ dependence that the
process of energy dissipation assumed in Kanatani's theory is not
valid in the present situation of the simulation.  Although, at
$\nu=0.8$, the effect of a multi-contact of the particle may be
significant, the velocity difference at the contact point would
not be  completely sustained during the contact.

\begin{figure}
\centerline{
\includegraphics[width=0.49\textwidth]{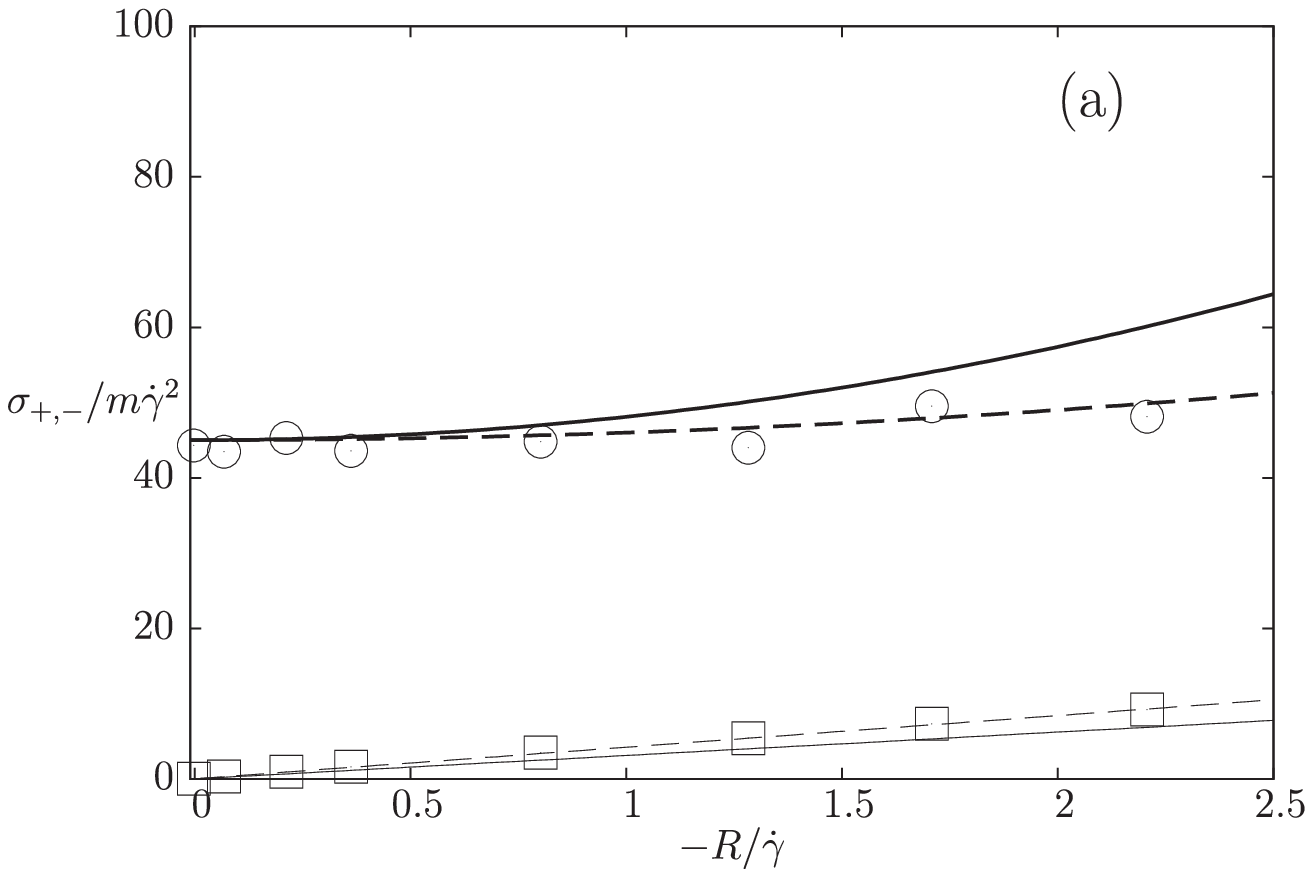}
\includegraphics[width=0.49\textwidth]{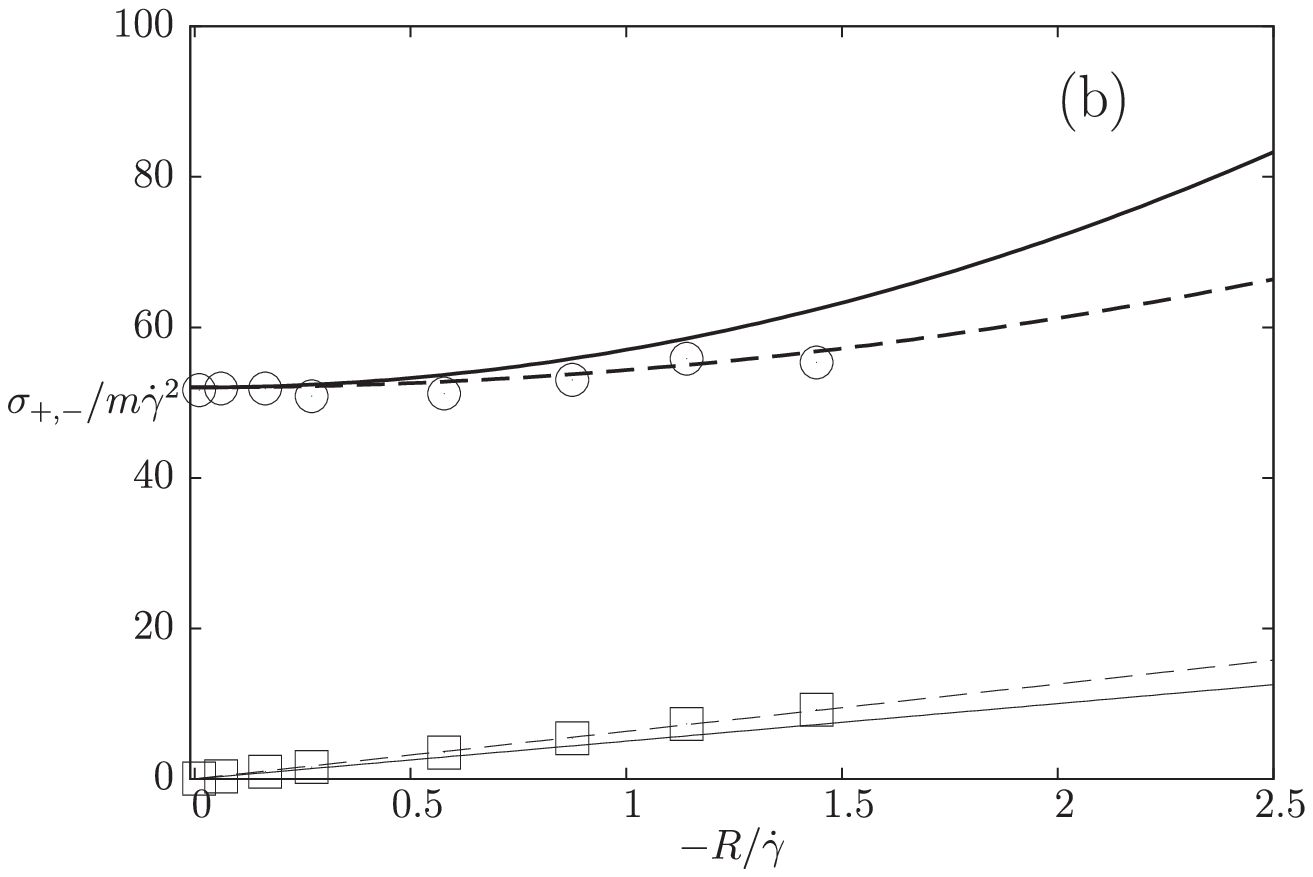}
}
\caption{\label{fig:sigmadecompose} Normalized stresses
$\sigma_\pm/m\dot\gamma^2$ as functions of $-R/\dot\gamma$
in the simulations and those based on
the decomposition by Kanatani (\ref{eq:sigmagammaR}) for
$\mu=0.3$ [(a)] and $\mu=0.8$ [(b)].  $\nu=0.8$ for both figures.
Circle and square symbols denote normalized $\sigma_+$ and $\sigma_-$
in the simulation, respectively.  Thick and thin dashed lines are
the fitted lines (\ref{eq:sigmafit1}), (\ref{eq:sigmafit2})
to $\sigma_+$ and $\sigma_-$, respectively.
Thick and thin solid lines indicate the decompositions
(\ref{eq:decompose1}), (\ref{eq:decompose2}) for
$\sigma_+$ and $\sigma_-$, respectively.}
\end{figure}
Now, let us examine the choice (\ref{eq:choice}) of
the constitutive equations in Kanatani's theory.
In order to focus on the examination of the choice
(\ref{eq:choice}), let us use $\Phi$ obtained in the simulation
rather than $\Phi$ of the estimates in Kanatani's theory.
The form (\ref{eq:phi_gammaR}) with (\ref{eq:sigma_scaling}) implies
that the degree of homogeneity $\zeta$ is $3$.
The choice (\ref{eq:choice}) reads, in the present context, as follows:
\be
\sigma_+=\frac{1}{3}\frac{\pd\Phi}{\pd\dot\gamma}\,,\qquad
\sigma_-=\frac{1}{6}\frac{\pd\Phi}{\pd R}\,.
\label{eq:sigmagammaR}
\ee
The normalized dissipation function
$\tilde\Phi:=\Phi/m\dot\gamma^3$ in the simulation
can be fitted by a polynomial of $r:=R/\dot\gamma$,
\be
\tilde\Phi=A + B r^2,
\label{eq:Phifit}
\ee
with
\begin{align}
A\simeq 45,\quad B\simeq 9.4, \qquad (\mu=0.3),
\label{eq:Phifit1}\\
A\simeq 52,\quad B\simeq 15, \qquad (\mu=0.8),
\label{eq:Phifit2}
\end{align}
(see figure~\ref{fig:Phi}).
Then, the choice (\ref{eq:sigmagammaR}) implies that
\be
\tilde\sigma_+=A_+ + B_+ r^2, \qquad
\tilde\sigma_-=A_- r,
\label{eq:sigmapoly}
\ee
with
\be
A_+= A,\quad B_+=\frac{B}{3}\,, \qquad A_-= \frac{B}{3}\,.
\label{eq:ABpm}
\ee
From (\ref{eq:Phifit1}), (\ref{eq:Phifit2}) and (\ref{eq:ABpm}),
one obtains
\begin{align}
&A_+\simeq 45,\quad B_+\simeq 3.1,\quad
A_-\simeq 3.1,\quad (\mu=0.3)
\label{eq:decompose1}
,\\
&A_+\simeq 52, \quad B_+\simeq 5.0,\quad
A_-\simeq 5.0, \quad (\mu=0.8).
\label{eq:decompose2}
\end{align}
On the other hand, we obtain
\begin{align}
&A_+\simeq 45, \quad B_+\simeq 1.0,\quad A_-\simeq 4.2,
\quad (\mu=0.3),
\label{eq:sigmafit1}\\
&A_+\simeq 52, \quad B_+\simeq 2.3, \quad A_-\simeq 6.3,
\quad (\mu=0.8),
\label{eq:sigmafit2}
\end{align}
by fitting the function of the form (\ref{eq:sigmapoly}) directly
to the simulation data.  One can see that they disagree.
Especially, in the decomposition based on (\ref{eq:sigmagammaR}),
we have $B_+=A_-$ according to (\ref{eq:ABpm}). However, this is
not the case in the simulation. The disagreement can be also
checked in figure~\ref{fig:sigmadecompose}; the normalized
stresses $\tilde\sigma_\pm=\sigma_\pm/m\dot\gamma^2$
of~(\ref{eq:sigmapoly}) with~(\ref{eq:decompose1})
and~(\ref{eq:decompose2}) are plotted together with the normalized
stresses directly measured in the numerical simulation and their
fits~(\ref{eq:sigmapoly}) with~(\ref{eq:sigmafit1})
and~(\ref{eq:sigmafit2}). Thus, the choice~(\ref{eq:choice}) of
the constitutive equations made by Kanatani is not appropriate in
the present situation of the simulation.

\section{Discussion}
\label{sec:discussion}
We confirmed in section~\ref{sec:comparisonkinetictheory} that
the constitutive equation for $\sigma_+$
in the kinetic theory by Lun is in good agreement with the simulation
results for a small area fraction $\nu=0.1$.  But the agreement
for $\sigma_-$ is not as good as $\sigma_+$.
Since $\sigma_-$ is sensitive to $\beta$, this may be due to the
oversimplification of modeling $\beta$ by a constant in the Lun's
theory.  As $\nu$ increases, discrepancy
between the kinetic theory and the simulation results
on both $\sigma_+$ and $\sigma_-$  increases.
However, $\sigma_-$ in the kinetic theory can be
formally fitted to the simulation results by introducing the effective
coefficient of roughness $\beta_{\rm eff}$ which is substantially
smaller than the actual averaged beta $\bar\beta$ in the simulation.
This implies that, although the particle-pairs almost get stuck
($\bar\beta\sim 0$) in the tangential direction during the collision,
the macroscopic field feels as if the particles were slipping
($\beta_{\rm eff} \sim -1$).  A possible scenario is that the collective
motions of some stuck particles affect the macroscopic field as
motions of virtual particles with renormalized mass $m$, radius $a$,
coefficient of restitution $e$, and $\beta$.
In the present study, we renormalized $\beta$ with fixed
$m$, $a$ and $e$.  Then $\beta$ is renormalized to reduce its value.


We have seen in section~\ref{sec:comparisonkanatanitheory}
that Kanatani's theory is not applicable to the present situation
of simulation ($\nu=0.8$) in the following sense;
(i) the kinetic friction coefficient $\mu$ dependence of the dissipation
function $\Phi$ and
(ii) the choice (\ref{eq:choice}) made for determining the constitutive
equations from $\Phi$, are in disagreement with those in the simulations.

To summarize, there is a regime of relatively dense ($\nu\sim 0.8$
in the present simulation)
granular flows in the form of micropolar fluid, where physical pictures based
on neither kinetic theories nor Kanatani's theory are adequate.
In this regime, a new microscopic description of particle interaction,
that is different from mutually independent short-time collisions in
the kinetic theory or long-time contacts with sustained velocity
difference in Kanatani's theory, is necessary.
As we have seen in sections~~\ref{sec:setting} and
\ref{sec:comparisonkinetictheory}, there are considerable
changes in the interfering number $n_{\rm int}$,
$\bar\beta(\vartheta)$ and $P(\vartheta)$ when $\nu$
increases from $0.7$ to $0.8$.  These changes suggest the significance
of the effect of $n$-particle interactions with $n>2$ in this regime.
It would be a future study to see whether the effect is included in
the kinetic theory with renormalized parameters as we have preliminary
analyzed or in some variant of Kanatani's theory with an appropriate
dissipation function and its decomposition.

\section*{Acknowledgements}
We are grateful to the anonymous referee for
his/her enlightening comments and notification of several important
papers, which led to significant expansion and improvement of the paper.

\newpage

\newpage

\ukrainianpart

\title{Матеріальні рівняння для гранульованого потоку  з однорідним середнім зсувом і спіновими полями
}

  \author{К. Такечі, К. Йошіда, Т. Аріміцу}
  \address{Вища школа фундаментальних  та прикладних наук, університет м. Цукуба, Ібаракі, Японія}

\date{Отримано 12 травня 2010 р., в остаточному вигляді 22 грудня 2010 р.}

\makeukrtitle

\begin{abstract}
\tolerance=3000%
Для того, щоб виокремити матеріальні рівняння для системи були здійснені числові симуляції двовимірних гранульованих потоків під дією однорідного зсуву і зовнішнього крутильного моменту.
Результат чисельних симуляцій проаналізовано на основі  моделі мікрополярного плину. В симуляціях реалізується поле однорідного середнього зсуву, яке не є підпорядковане полю вихоровості. Оцінки напружеь, зроблені на основі кінетичної теорії Люна [Lun, J. Fluid Mech. {\bf 233}(1991) 539], добре узгоджуються з результатами симуляцій в області низьких часток
$\nu=0.1$, але узгодження погіршується, якщо ця величина зростає.  Проте, оцінки, зроблені в  кінетичній теорії можуть бути підігнані до результатів симуляцій аж до  $\nu=0.7$ шляхом ренормалізації коефіцінта шорсткості.  Для відносно густого гранульованого потоку  ($\nu=0.8$),
результати симуляцій також порівнюються з теорією Канатані [Kanatani, Int. J. Eng. Sci {\bf 17}(1979) 419]. Знайдено, що дисипативна функція і її декомпозиція в матеріальні рівняння в теорії Канатані не узгоджується з результатами симуляцій.
\keywords гранульований потік, матеріальні рівняння, мікрополярний плин, кінетичне рівняння
\end{abstract}

\end{document}